\documentclass[twocolumn,amsmath,amssymb,aps,showkeys]{revtex4-1}
\usepackage{graphicx}
\usepackage{amsmath,amssymb}
\usepackage{mathrsfs}
\usepackage{color}
\usepackage{afterpage}
\usepackage[version=3]{mhchem}
\usepackage{natbib}
\usepackage{soul}
\usepackage[caption=false]{subfig}
\usepackage{array}
\usepackage{multirow}
\usepackage{float}
\usepackage[colorlinks, citecolor=blue,urlcolor=blue, linkcolor=blue, bookmarks=false]{hyperref}
\hypersetup{colorlinks=true , citecolor=blue, urlcolor=blue, linkcolor=blue}

\begin{document}
\title{Manipulation of valley and spin properties in two-dimensional Janus WSiGeZ$_4$ (Z=N, P, As) through symmetry control}

\author{Sajjan Sheoran\footnote{phz198687@physics.iitd.ac.in}, Ankita Phutela, Ruman Moullik, and Saswata Bhattacharya\footnote{saswata@physics.iitd.ac.in}} 
\affiliation{Department of Physics, Indian Institute of Technology Delhi, New Delhi 110016, India}

\begin{abstract}
	\noindent
	 
	A class of septuple-atomic-layer two-dimensional (2D) materials, MA$_2$Z$_4$, is sought as an alternative to 2D hexagonal transition metal dichalcogenides in the field of valleytronics and spintronics. In these materials, the structural symmetry can be varied by changing the stacking of its three parts in the monolayer. We show that in Janus monolayer WSiGeZ$_4$ (Z=N, P, As), the Berry curvature and Rashba effect are enhanced by modifying the stacking orders. Intrinsic electric field and composition of the $d$ orbitals play a dominant role in determining these properties. The vertical field lifts the spin degeneracy along in-plane direction, inducing the Rashba effect around $\Gamma$ point. The in-plane orbitals under the influence of in-plane electric field contribute to the Zeeman splitting of bands at K/K$^{\prime}$ points. Berry curvature is strengthened by up to 300\% compared to its ground state through symmetry control, along with a significant increment in the Rashba coefficient. Moreover, monolayer WSiGeP$_4$ and WSiGeAs$_4$ have multiple valleys, implying another valley-dimension. We construct a symmetry adapted \textit{\textbf{k.p}} Hamiltonian for the valleys and investigate the effect of strain and electric field on the band structure. The interesting spin-valley physics in monolayer WSiGeZ$_4$ suggests their exceptional potential for spintronics and valleytronics applications.
\end{abstract}
\maketitle

\section{Introduction}
The successful investigation of graphene has spurred the discovery of extraordinary properties and promising applications in the field of two-dimensional (2D) materials~\cite{geim2009graphene,novoselov2012roadmap,kane2005quantum}. Afterwards, numerous 2D materials such as silicene, germanene, boron nitride, transition metal dichalcogenides (TMD) and so on have been explored~\cite{molle2018silicene,ni2012tunable,chopra1995boron,manzeli20172d,xu2014spin,wang2018colloquium}. 2D materials with Janus structure possess different physical and chemical properties on each side due to mirror symmetry breaking, which make them exciting~\cite{liang2014rational,li2018recent}. The asymmetric crystal potential causes an intrinsic electric field in these Janus 2D materials, leading to spin-valley splitting and Rashba spin splitting when combined with spin-orbit coupling (SOC)~\cite{hu2018intrinsic,touski2021structural}. Additionally, Janus materials show great applications in photocatalytic water splitting, magnetic skyrmions, and out-of-plane piezoelectricity~\cite{singh2021mos,yuan2020intrinsic,dong2017large}. The experimental synthesis of MoSSe from MoSe$_2$ through the replacement of Se atom by S atom, further triggered the prediction of more Janus 2D materials~\cite{lu2017janus,jang2022growth}. Only TMD and metal monochalcogenides Janus materials have been extensively studied so far. Therefore, it is necessary to expand this family and probe into their physical properties to have a more in-depth understanding.

Recently, the MA$_2$Z$_4$ (M=Cr, Mo, W, V, Nb, Ta, Ti, Zr or Hf; A=Si or Ge; and Z=N, P, As) 2D family of the septuple-atomic layer has been proposed~\cite{wang2021intercalated}. Particularly, 2D van der Waals (vdW) layered MoSi$_2$N$_4$ and WSi$_2$N$_4$ have been successfully fabricated by the chemical vapor deposition method~\cite{hong2020chemical}. It has been found that WA$_2$Z$_4$ materials have high tensile strength with excellent ambient stability. Subsequently, several predictions of these Janus 2D materials and their excellent physical properties have been carried out. This includes the quantum spin Hall state in SrAlGaTe$_4$~\cite{guo2021piezoelectric}, high electron mobility, efficient photocatalysis, Rashba spin splitting, out-of-plane piezoelectricity in the MoSiGeN$_4$~\cite{rezavand2022electronic, guo2021predicted, hussain2022emergence} and intrinsic ferromagnetism in the VSiGeN$_4$~\cite{dey2022intrinsic,guo2022strain}. Therefore, this class of materials is expected to provide intriguing physical phenomena, particularly in valleytronics and spintronics.

Nowadays, spin and valley are considered as extra degrees of freedom, which are distinct from the electronic charge and provide an interesting avenue in semiconductor devices~\cite{xiao2012coupled,islam2021tunable}. This draws a great deal of attention to the number of emergent quantum phenomena such as valley Hall effect and valley-dependent orbital magnetic moment. 
The broken inversion symmetry is compulsory for valleytronics, since Berry curvature vanishes in centrosymmetric materials~\cite{PhysRevLett.123.196403}. Also, strong SOC in inversion asymmetric materials induces Rashba spin splitting, which is a critical ingredient in spin field-effect transistors (SFETs)~\cite{hurand2015field,D1MA00912E,PhysRevMaterials.6.094602}. The charge carriers with Rashba SOC experience a momentum-dependent effective magnetic field, hence a spin-dependent velocity correction term arises~\cite{D1MA00912E}. Thus, the Rashba effect is extensively studied for SFETs and spin-to-charge interconversion. The spin-valley locking and the Rashba effect arising from the SOC are bringing tremendous attention to the flourishing field of valleytronics and spintronics. These effects have been noticed in graphene~\cite{xiao2007valley}, TMDs~\cite{xiao2012coupled}, $h$-MN (M=Nb, Ta)~\cite{ahammed2022valley} and bismuth thin films~\cite{jin2021enhanced}. However, the scarcity of 2D materials showing these phenomena with appreciable spin-valley polarization and significant Rashba effect holds back the field of spin valleytronics.

In this article, we have investigated the effect of stacking of A-Z layers to modulate the electronic properties, focusing on their spintronics and valleytronics aspects. We have taken the unexplored Janus WSiGeZ$_4$ as an example to study these effects. The first-principles calculations predict that these materials have Dirac-type valleys located at the corners of Brillouin zone (BZ), connected by time-reversal symmetry. The valleys show contrasting physics, i.e., spin polarization, optical circular dichroism and Berry curvature. The broken surface inversion symmetry induces Rashba spin splitting at the center of the BZ. The findings imply that symmetry control enhances Berry curvature by more than 300\%. Furthermore, Rashba splitting increases significantly by modifying the out-of-plane asymmetry. The underlying mechanism behind the unconventional enhancement is duly investigated through microscopic orbital contribution and macroscopic charge transfer. This method is more effective compared to the strain and electric field, in modulating the Berry curvature and Rashba splitting. The high carrier mobilities and tunable electronic properties suggest their great potential in spintronics and valleytronics.

\section{Computational Methods}
Density functional theory (DFT) calculations have been carried out to study the electronic properties as implemented in Vienna \textit{ab initio} simulation package (VASP)~\cite{kresse1996efficient}. Projector augmented waves (PAW) with cutoff energy of 550 eV are employed as a basis set~\cite{blochl1994projector,kresse1999ultrasoft}. The Perdew-Burke-Ernzerhof (PBE) parametrization at the level of generalized gradient approximation is used as the exchange-correlation functional~\cite{perdew1996generalized}. The structures are initially relaxed until the force on each atom is smaller than 1 meV/\AA. A vacuum spacing of 18 {\AA} vertical to the layers is maintained to avoid spurious interactions between periodic images. The energy convergence threshold is set to $10^{-6}$ eV. BZ sampling is done using \textit{k}-point mesh with a separation of 0.02 \AA$^{-1}$. Phonon spectra are calculated using the density functional perturbation theory as implemented in VASP alongside PHONOPY~\cite{togo2015first}. Berry curvatures and spin textures are obtained from the maximally localized Wannier functions as implemented in the Wannier90~\cite{mostofi2014updated}. Band gaps are also calculated using the hybrid Heyd-Scuseria Ernzerhof (HSE06)~\cite{heyd2003hybrid} functional and G$_0$W$_0$@PBE method~\cite{hedin1965new,hybertsen1985first}. To appraise the optical spectra, many-body perturbation theory calculations within the framework of G$_0$W$_0$ and BSE~\cite{albrecht1998ab} are performed.

\begin{figure}[ht]
	\includegraphics[width=8.5cm]{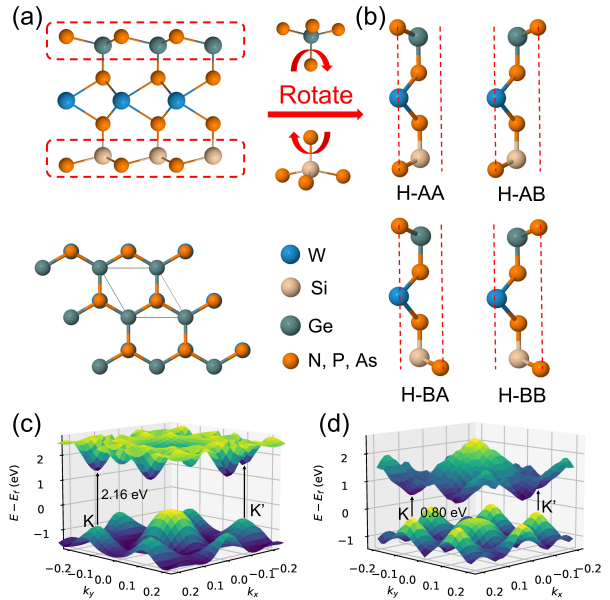}
	\caption{(a) Side and top views of lattice structure corresponding to monolayer H-BB WSiGeZ$_4$. (b) Side views of various stacking orders for three constituent parts in the monolayer WSiGeZ$_4$. The 3D view of the top valence band and bottom conduction band in monolayer (c) H-BB WSiGeN$_4$ and (d) H-AA WSiGeP$_4$.}
	\label{p1}
\end{figure}
\begin{table*}
	\begin{center}
		\caption{ Optimized lattice constant ($a$) for the unit cell of WSiGeZ$_4$ monolayer. The calculated formation energy per atom ($E_{for}$), dipole moment per unit cell ($P_y$, $P_z$), band gaps in presence of SOC using PBE ($E_g^{\textrm{PBE}}$) and HSE06 ($E_g^{\textrm{HSE}}$). The calculated band gaps using $\textrm{G}_0\textrm{W}_0$ performed on top of PBE functional with inclusion of SOC ($E_g^{\textrm{GW@PBE}}$).}
		\begin{tabular}{p{1.6cm} p{1.2cm} p{1.2cm} p{1.2cm} p{1.2cm} p{1.2cm} p{1.2cm} p{1.2cm}p{1.6cm}}
			\hline
			\hline
			& Phase  & $a$  & $E_{for}$ & $P_y$ & $P_z$  & $E_g^{\textrm{PBE}}$ &$E_g^{\textrm{HSE}}$  &$E_g^{\textrm{GW@PBE}}$    \\
			& &({\AA}) & (eV) & (e{\AA}) & (e\AA) &  (eV) & (eV) & (eV)  \\\hline  
			WSiGeN$_4$ & H-AA &  2.929 & -0.639 & 1.710 & 0.024 & 2.01 & 2.48 & 3.27  \\
			& H-AB &  2.931 & -0.643 & -0.930 & 0.021 & 1.88 & 2.36 &3.11   \\
			& H-BA &  2.934 & -0.651 & -2.488 & 0.028 & 1.83 & 2.31 & 3.03  \\
			& H-BB &  2.940 &  -0.658& 1.716 & 0.026 & 1.66 & 2.15 & 2.87 \\
			WSiGeP$_4$ & H-AA &  3.452 & -0.279 & 2.016 & 0.042 & 0.56 & 0.58 & 1.19  \\
			& H-AB &  3.459 & -0.259 & -1.085 & 0.032 & 0.46 & 0.51 & 1.06  \\
			& H-BA &  3.462 &  -0.271& -2.989 & 0.058 & 0.39 & 0.47 & 0.97  \\
			& H-BB &  3.468 &  -0.262& 2.031 & 0.043 &  0.25 & 0.42 & 0.75  \\
			WSiGeAs$_4$& H-AA &  3.594 & -0.113 & 2.103 & 0.096 & 0.41 & 0.45 & 0.89  \\
			& H-AB &  3.599 & -0.072 & -1.204 & 0.084 & 0.33 & 0.39 & 0.82  \\
			& H-BA &  3.601 & -0.086& -3.104 & 0.124 & 0.28 & 0.36 & 0.76 \\
			& H-BB &  3.605 & -0.053& 2.118 & 0.101 &  0.18 &  0.29 & 0.58  \\
			\hline\hline
		\end{tabular}
		\label{T1}
	\end{center}
\end{table*}
\section{Results and Discussion}
\subsection{Crystal structure and electronic properties}
WA$_2$Z$_4$ structure, built by septuple atomic layers in the sequence Z-A-Z-W-Z-A-Z having $D_{3h}$ point group symmetry, can be regarded as one layer of WZ$_2$ sandwiched by two layers of A-Z~\cite{wang2021intercalated}. The intermediate WZ$_2$ is structurally similar to 1H-TMD~\cite{wang2021intercalated,sheoran2022coupled,sheoran2023coupled}. The WA$_2$Z$_4$ structure breaks inversion symmetry and preserves horizontal mirror symmetry ($\sigma_h$). When they form the Janus structure by replacing A atom with a different same group atom, $\sigma_h$ is broken, leading to inversion symmetry breaking with $C_{3v}$ symmetric point group~\cite{rezavand2022electronic, guo2021predicted}.
Figure~\ref{p1}(a) shows the fully optimized crystal structure of the WSiGeZ$_4$. Theoretically, several intermediate structures are obtained by varying the stacking of the intermediate layer of 1H-WZ$_2$ and two A-Z layers~\cite{hong2020chemical,wang2021intercalated,zhou2021structural,kang2021second,wang2022two,ren2022two,guo2021structure}. As shown in Fig.~\ref{p1}(b), we have considered the four structures (H-AA, H-AB, H-BA, and H-BB) depending upon the stacking order of the three constituent (Z-Si, Z-W-Z, and Ge-Z) monolayers. As shown in Table~\ref{T1}, the lattice parameters of H-AA, H-AB, H-BA and H-BB are only slightly different. This is because of same coordination environment and similar valence electron states. The lattice constant of Janus WSiGeZ$_4$ monolayer is approximately the average of WSi$_2$Z$_4$ and WGe$_2$Z$_4$. It is evident that the lattice constant increases with the increasing atomic number of the pnictogen.

We have obtained the formation energy per atom ($E_{for}$) to confirm the energy feasibility, using the expression
\begin{equation}
	E_{for}=\frac{E_{tot}-(n_WE_W+n_{Si}E_{Si}+n_{Ge}E_{Ge}+n_ZE_Z)}{(n_W+n_{Si}+n_{Ge}+n_Z)}
\end{equation}
where $E_{tot}$ is the total ground state energy of the WSiGeZ$_4$ monolayer, $E_W$, $E_{Si}$, $E_{Ge}$, and $E_Z$ are the chemical potentials of W, Si, Ge, and Z atoms, respectively. $n_W$, $n_{Si}$, $n_{Ge}$, and $n_Z$ are the number of W, Si Ge, and Z atoms in the unit cell, respectively. As clear from Table~\ref{T1}, all the considered phases are energetically feasible. Nevertheless, the ground state configurations are different. For instance, the H-BB phase is most favorable for WSiGeN$_4$, but WSiGeP$_4$ and WSiGeAs$_4$ are most stable in the H-AA phase. The dynamical stability of these structures is verified by calculating phonon spectra (see section I of Supplemental Material (SM)). The absence of imaginary phonon mode in the entire BZ confirms the stability of phases H-AA, H-AB, H-BA, and H-BB. Furthermore, we have calculated the electric polarization using the Berry-phase method~\cite{king1993theory,spaldin2012beginner}. There are two main factors contributing to the origin of vertical polarization. The first one is the asymmetry arising from the replacement of one Si by Ge atom, and another is the asymmetry arising from the rotation of Si-Z and Ge-Z layers (see Fig.~\ref{p1}(b)). Table~\ref{T1} summarizes the calculated electric dipole moment per unit cell. 

\begin{figure}[ht]
	\includegraphics[width=8 cm]{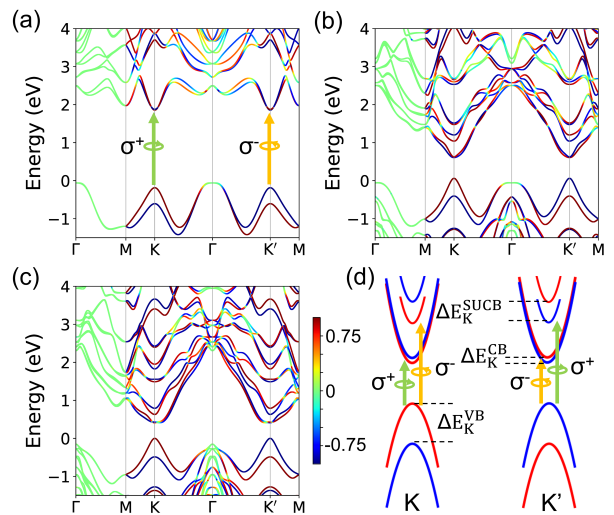}
	\caption{The spin resolved band structure of monolayer (a) H-BB WSiGeN$_4$, (b) H-AA WSiGeP$_4$ and (c) H-AA WSiGeAs$_4$ calculated using PBE functional with inclusion of SOC. (d) Schematic of valley selective excitation. Length of arrows and circles depict the energy and chirality of incident photon, respectively. The red and blue colors denotes the spin-up and spin-down states, respectively. }
	\label{p2}
\end{figure}
\begin{figure*}[ht]
	\includegraphics[width=12cm]{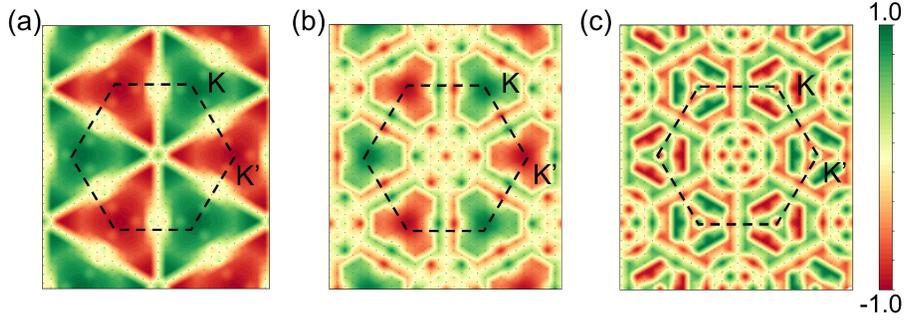}
	\caption{The degree of circular polarization for transition between VB to CB in monolayer (a) H-BB WSiGeN$_4$ and (b) H-AA WSiGeP$_4$ mapped over the BZ. (c) The circular polarization for transition between VB to SUCB in monolayer H-AA WSiGeP$_4$. The black dotted lines represents the first BZ.}
	\label{p3}
\end{figure*}
The three-dimensional band structures of the monolayer H-BB WSiGeN$_4$ and H-AA WSiGeP$_4$ are shown in Figs. \ref{p1}(c) and \ref{p1}(d), respectively (see section II of SM for WSiGeAs$_4$). The monolayer H-BB WSiGeN$_4$ has an indirect band gap with valence band maximum (VBM) and conduction band minimum (CBM) at $\Gamma$ and K/K$^{\prime}$ points, respectively. The monolayer H-AA WSiGeP$_4$, and WSiGeAs$_4$ are the direct band gap semiconductors with band edges located at K/K$^{\prime}$ points. The monolayer WSiGeN$_4$, WSiGeP$_4$ and WSiGeAs$_4$ have valley-like peaks in the vicinity of K and K$^{\prime}$ points. The two valleys at K and K$^{\prime}$ are inequivalent due to structural asymmetry and time-reversal symmetry, although they have degenerate energy, as discussed in the coming sections.
\subsection{Valley physics}
To investigate the valleys in the monolayer WSiGeZ$_4$, band structures with SOC are calculated as shown in Figs.~\ref{p2}(a), \ref{p2}(b) and \ref{p2}(c). We have observed that monolayer H-BB WSiGeN$_4$ is still an indirect band semiconductor and monolayer H-AA WSiGeP$_4$ and WSiGeAs$_4$ remain direct band semiconductors. Band gaps obtained using PBE functional are listed in Table~\ref{T1}. We have also used the more accurate HSE06 functional and G$_0$W$_0$ approach to validate the band structure results. The band structures are shown in section III and IV of SM, and the band gaps are compared in Table~\ref{T1}. We have found that apart from the band gap, all the band features are similar irrespective of the choice of functional. The band gaps increase for the HSE06 and G$_0$W$_0$ methods due to the inclusion of exact exchange and self-energy, respectively. As seen in Figs.~\ref{p2}(a), \ref{p2}(b) and \ref{p2}(c), SOC mainly causes spin splitting of the degenerate bands. The spin splitting is a direct consequence of the inversion symmetry breaking, which lifts the spin degeneracy at each generic \textit{k}-point. Since the time-reversal symmetry is preserved, the spin polarization at K and K$^{\prime}$ points are opposite because the time-reversal operation connects the K and K$^{\prime}$ points. Additionally, properties that are odd under time-reversal operation will have opposite natures at K and K$^{\prime}$ points. This concludes the existence of a valley degree of freedom.

The circular polarization for optical transition between the pair of states $|n\textbf{\textit{k}}\rangle $ and $|m\textbf{\textit{k}}\rangle$ is given by~\cite{yang2021valley}
\begin{equation}
	\eta(\textbf{\textit{k}})=\frac{|M_+|^2-|M_-|^2}{|M_+|^2+|M_-|^2}
	\label{ecp}
\end{equation}
where $M_{\pm}$ is the optical transition matrix element for the incident light with $\sigma_{\pm}$ polarization. The $M_{\pm}$ can be expressed as $M_{\pm}=\frac{1}{\sqrt{2}}(M_x\pm M_y)$, where $M_{x/y}=m_e\langle m\textbf{\textit{k}}|v_{x/y}|n\textbf{\textit{k}} \rangle $. $\eta=+1$ and $\eta=-1$ represent the optical absorption of only right hand circularly polarized (RHCP) and left hand circularly polarized (LHCP) light, respectively. Figure~\ref{p3}(a) shows the degree of circular polarization of excitation from first occupied valence band (VB) to first unoccupied conduction band (CB) in the case of monolayer H-BB WSiGeN$_4$. As seen, the circular dichroism is perfectly valley selective. The absorption of RHCP and LHCP photon will take place at K and K$^{\prime}$ points, respectively. Therefore, the circularly polarized optical pumping can control the electron and hole population of either valley. In the case of the monolayer H-BB WSiGeN$_4$, the direct band gap at K/K$^{\prime}$ point is 2.16 eV. Therefore, either valley can be selectively excited by the circularly polarized photon with the energy around 2.16 eV.

In case of monolayer H-AA WSiGeP$_4$ and WSiGeAs$_4$, the picture remains the same. However, there are no additional states apart from K/K$^{\prime}$ near the CBM and VBM, thus, there will be no scattering states. Figure~\ref{p3}(b) shows the circular polarization for optical transition between VB to CB in the case of monolayer H-AA WSiGeP$_4$. The valleys are perfect and show complete valley contrasting physics, i.e., selective valley excitation. The valley selective optical transition occurs from VB to CB when the energy of the circularly polarized incident photon is around 0.56 and 0.41 eV for monolayer H-AA WSiGeP$_4$ and WSiGeAs$_4$, respectively. Furthermore, there exist additional valleys at the second unoccupied conduction band (SUCB) near the CBM (see Figs.~\ref{p2}(b) and \ref{p2}(c)). Also, the spin polarization for the SUCB at K and K$^{\prime}$ points is opposite. Therefore, if the energy of the incident photon is greater than the direct band gap, then the transition from VBM to SUCM takes place (see Fig.~\ref{p2}(d)). The circular polarization from the VB to SUCM is also valley specified, as shown in Fig.~\ref{p3}(c). The LHCP and RHCP from VBM to SUCM will be absorbed at K and K$^{\prime}$ valleys, respectively. Note that the chirality of light for the transition taking place from VB to CB or VB to SUCB is opposite in the same valley (see Fig.~\ref{p2}(d)). Therefore, we can selectively pump an electron from VB to CB or SUCB by using optically selective light. This shows that SUCB in monolayer H-AA WSiGeP$_4$ and WSiGeAs$_4$ add another dimension to the valley physics.

\begin{figure}
	\includegraphics[width=6cm]{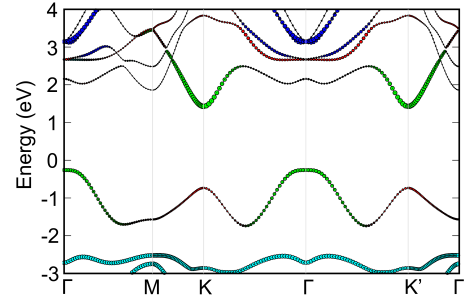}
	\caption{Orbital projected band structure of monolayer H-BB WSiGeN$_4$ without SOC. Fermi level is set to zero in the energy axis. The fatness of band structure corresponds to W-${d_{z^2}}$ (green), W-${d_{x^2-y^2}}$ and W-${d_{xy}}$ (red), W-${d_{yz}}$ and W-${d_{xz}}$ (blue), and N-$p$ (cyan) orbitals.}
	\label{p4}
\end{figure}

The low power \textit{\textbf{k.p}} Hamiltonian including the orbital and spin degrees of freedom explains the valley splitting around $C_3$ symmetric K/K$^{\prime}$ points. Figure~\ref{p4} shows the orbital-projected band structure of monolayer H-BB WSiGeN$_4$. The VB and SUCB states mainly consist of in-plane orbitals ($d_{xy}$, $d_{x^2-y^2}$). Whereas, the out-of-plane orbital ($d_{z^2}$) mainly contribute to the CBM. Similar contributions are also observed for monolayer H-AA WSiGeP$_4$ and WSiGeAs$_4$. The \textit{\textbf{k.p}} model Hamiltonian employing the basis $|d_{xy}$-$\tau$$i$$d_{x^2-y^2}\rangle \otimes |s_z\rangle$, $|d_{z^2} \rangle \otimes |s_z\rangle$ and $|d_{xy}$+$\tau$$i$$d_{x^2-y^2}\rangle \otimes|s_z\rangle$ and containing the terms up to linear in \textbf{\textit{k}} is given by~\cite{xiao2012coupled, li2020valley, yang2021valley}
\begin{equation}
	H_\tau =\begin{pmatrix}
		\epsilon_a+\tau \lambda_a S_z & \gamma_2 k^- & \gamma_1k^+\\
		\gamma_2k^+ & \epsilon_c & \gamma_3k^- \\
		\gamma_1k^+ & \gamma_3k^+ & \epsilon_v + \tau \lambda_v S_z
	\end{pmatrix}
\end{equation}
where $k^{\pm}=k_x\pm \tau i k_y$ and $k_x$ or $k_y$ is measured from K and K$^{\prime}$ point. The $\tau=\pm 1$ is the valley index and $\gamma_i$ are the optical matrix elements. The $\epsilon_c$, $\epsilon_v$, and $\epsilon_a$ are the energy levels at K/K$^{\prime}$ point. $\Delta=\epsilon_c-\epsilon_v$ is the direct band gap. $2\lambda_a$ ($\Delta E_K^{SUCB}$) and $2\lambda_v$ ($\Delta E_K^{VB}$) represent the spin splitting at the bottom of SUCB and the top of VB (see Fig.~\ref{p2}(d)). For the case of monolayer H-BB WSiGeN$_4$, SUCB is far away from the CBM. Therefore, the two band \textit{\textbf{k.p}} Hamiltonian is sufficient, which reads as~\cite{xiao2012coupled}
\begin{equation}
	H_\tau =\begin{pmatrix}
		\epsilon_c & \gamma_3k^- \\
		\gamma_3k^+ & \epsilon_v + \tau \lambda_v S_z
	\end{pmatrix}
   \label{e4}
\end{equation}
\begin{table}
	\begin{center}
		\caption{The observed spin splitting of VB ($\Delta E_K^{VB}$), CB ($\Delta E_K^{CB}$), SUCB ($\Delta E_K^{SUCB}$) and the Berry curvature ($\Omega_K$) at K point.}
		\begin{tabular}{p{1.6cm} p{1.2cm} p{1.2cm} p{1.2cm} p{1.4cm} p{1.0cm} }
			\hline
			\hline
			& Phase  & $\Delta E_K^{VB}$  & $\Delta E_K^{CB}$ & $\Delta E_K^{SUCB}$ & $\Omega_K$     \\
			& &(meV) & (meV) & (meV) & (\AA$^2$)  \\\hline  
			WSiGeN$_4$ & H-AA  & 415 & 3 & - & 16.1    \\
			& H-AB  & 415 & 4 & - & 17.2    \\
			& H-BA  & 417 & 9 & - & 17.5    \\
			& H-BB  & 418 & 12& - & 18.1    \\
			WSiGeP$_4$ & H-AA  & 431 & 6 & 184 & 75  \\
			& H-AB  & 426 & 4 & 186 & 133  \\
			& H-BA  & 435 & 6 & 174 & 161  \\
			& H-BB  & 436 &  9 & 184 & 335  \\
			WSiGeAs$_4$& H-AA  & 480 & 18 & 164 & 108 \\
			& H-AB  & 481 & 25 & 217 & 151  \\
			& H-BA  & 492 & 26 & 180 & 205 \\
			& H-BB  & 502 & 28 & 236 & 411 \\
			
			\hline\hline
		\end{tabular}
		\label{T2}
	\end{center}
	
\end{table}
\begin{figure}
	\includegraphics[width=8.5cm]{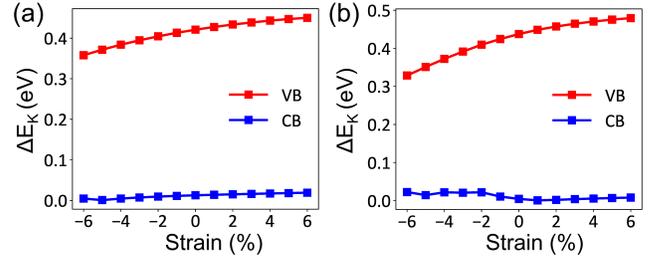}
	\caption{The variation of spin splitting of the VB and CB at K/K$^{\prime}$ valleys with respect to the biaxial strain in monolayer (a) H-BB WSiGeN$_4$ and (b) H-AA WSiGeP$_4$.}
	\label{p5}
\end{figure}
\begin{figure*}
	\includegraphics[width=12cm]{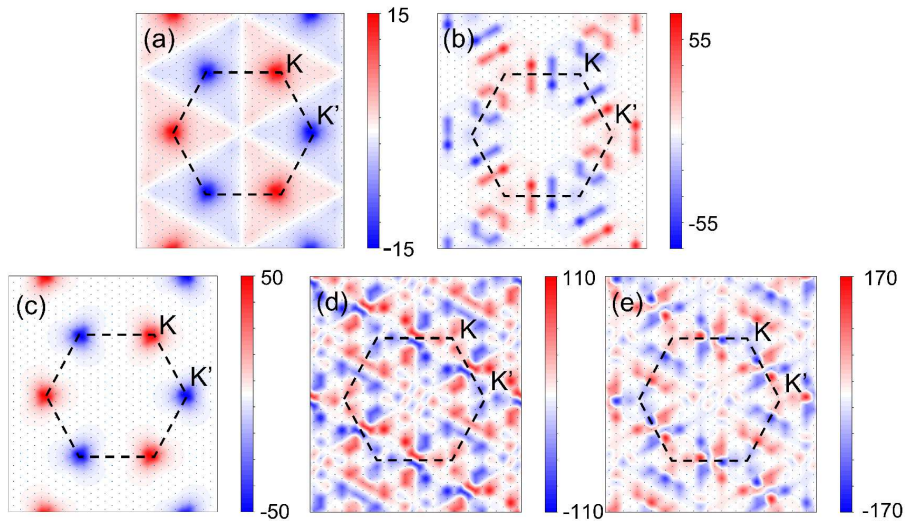}
	\caption{Berry curvature distribution over BZ for (a) VB, (b) CB in monolayer H-BB WSiGeN$_4$. Berry curvature distribution of (c) VB, (d) CB and (e) SUCB in monolayer H-AA WSiGeP$_4$. The black dotted lines denote the first BZ.}
	\label{p6}
\end{figure*}

To quantize the SOC effect, we have calculated the strength of the spin splitting at the top of the VB, the bottom of CB and SUCB (see Table~\ref{T2}). The valley splitting observed for the top VB and SUCB is significantly larger than the lowest CB. The W-$d_{xy}$ and W-$d_{x^2-y^2}$ orbitals under the influence of in-plane electric dipole moment produce spin splitting at the K point~\cite{PhysRevB.86.045437}. The CB is mainly contributed by W-$d_{z^2}$ and the contributions from W-$d_{xy}$ and W-$d_{x^2-y^2}$ are nearly zero. Therefore, spin splitting is predominant for the VB and the SUCB, and very tiny for the CB. The trend for spin splitting at valleys is observed as WSiGeN$_4$ $<$ WSiGeP$_4$$ <$ WSiGeAs$_4$, which is attributed to the increase in SOC for heavier elements.

Incommensurate lattices can bring about interfacial strain during support on a substrate or vdW heterostructures formation. Therefore, the role of strain is also addressed in this study. We have applied the biaxial compressive and tensile strain in the deformation range of 6\% to investigate the evolution of band edges. Figures~\ref{p5}(a), \ref{p5}(b) and \ref{p5}(c) show the variation of spin splitting at valleys of CB and VB for monolayer WSiGeZ$_4$. We see that the $\Delta E_K^{CB}$ is independent of the strain, whereas the $\Delta E_K^{VB}$ slightly increases (decreases) with the application of tensile (compressive strain).

\subsection{Berry curvature modulation}
To confirm the nonequivalence of valleys at K and K$^{\prime}$ points, we have studied the non-zero out-of-plane Berry curvatures. The velocity of the charge carriers in the presence of non-zero Berry curvature is given by
$\hbar v_n(\textbf{\textit{k}})=v_g-v_\perp$. Here, $v_g$ denotes the group velocity, which is defined as $v_g=\nabla_\textbf{\textit{k}}\epsilon_n(\textbf{\textit{k}})$, where $\epsilon_n(\textbf{\textit{k}})$ is the energy of $|n\textbf{\textit{k}}\rangle$ state. In addition to the group velocity, Berry curvature leads to a transverse velocity denoted by $v_\perp$. The transverse velocity is given by $v_\perp=-\frac{e}{\hbar}E\times\Omega_n(\textbf{\textit{k}})$, where E is the in-plane electric field and $\Omega(\textbf{\textit{k}})$ is the out-of-plane Berry curvature~\cite{yang2021valley}. For 2D systems, Berry curvature only has a $z$-component and for $|n\textbf{\textit{k}}\rangle$ state, it is expressed as~\cite{yang2021valley} 
\begin{equation}
	\Omega_n(\textbf{\textit{k}})=-2\textrm{Im}\sum_{n'\neq n} \frac { \langle n\textbf{\textit{k}}|v_x|n'\textbf{\textit{k}}\rangle \langle n'\textbf{\textit{k}}|v_y|n\textbf{\textit{k}}\rangle }{(\epsilon_{n'}-\epsilon_n)^2}
	\label{e5}
\end{equation}
where the summation is over all the occupied bands. $\epsilon_{n}$ and $\epsilon_{n'}$ are the energy levels for $|n\textbf{\textit{k}}\rangle$ and $|n'\textbf{\textit{k}}\rangle$ states, respectively. The time-reversal operation applies the opposite Berry curvature at time-reversal conjugate momenta ($\Omega_n(-K)=-\Omega_n(K)$). Figures~\ref{p6}(a) and \ref{p6}(b) show the Berry curvature of the VB and CB mapped in the first BZ for monolayer H-BB WSiGeN$_4$. Figures~\ref{p6}(c), \ref{p6}(d) and \ref{p6}(e) show the Berry curvature of monolayer H-AA WSiGeP$_4$ for VB, CB, and SUCB, respectively. Berry curvature is mostly confined around the valleys and has been found to be the highest at K and K$^{\prime}$ points. It has also been observed that the Berry curvature rapidly decreases to zero while moving away from the K and K$^{\prime}$ points. The opposite nature of Berry curvature near the K and K$^{\prime}$ points can distinguish the charge carriers at different valleys. The opposite Berry curvature leads to the spatial separation of charge carriers coupled with different valleys. Thus, the charge from different valleys will accumulate on opposite edges, leading to valley Hall effect. The time-reversal symmetry leads to the case of no net charge current at the linear order. However, there could exist a valley Hall effect after the electron doping. As for hole doping, one has to reach the energy level of K/K$^{\prime}$ point which is lower in energy compared to the $\Gamma$ point for monolayer WSiGeN$_4$. The monolayer WSiGeP$_4$ and WSiGeP$_4$ are more suitable for the case of hole doping, as the valence states lie at the K/K$^{\prime}$ point.

\begin{figure}[ht]
	\includegraphics[width=8.5cm]{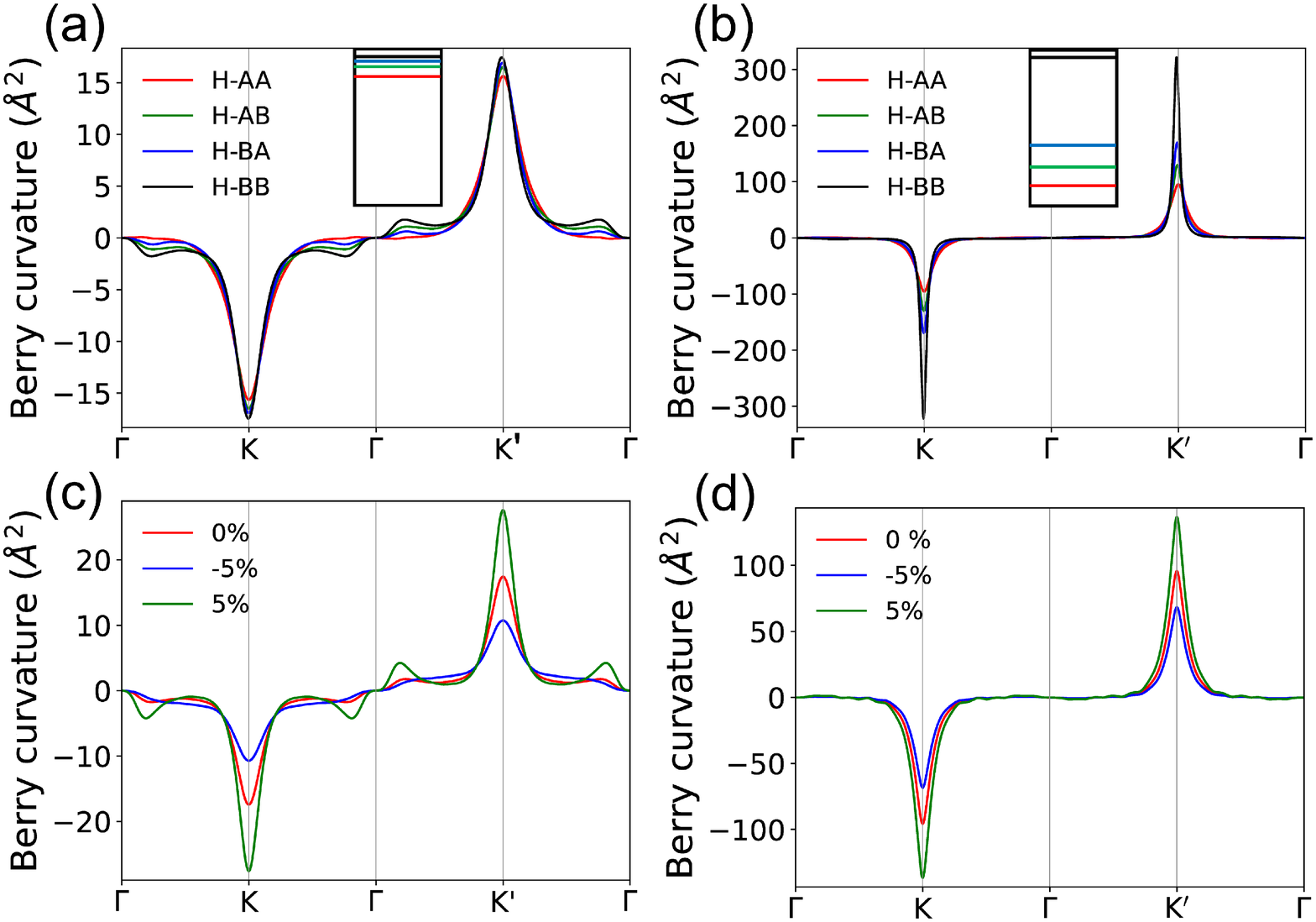}
	\caption{ Modulation of Berry curvature by varying stacking orders in monolayer (a) WSiGeN$_4$ and (b) WSiGeP$_4$ along the high symmetry line $\Gamma$-K-$\Gamma$-K$^{\prime}$-$\Gamma$. The insets in (a) and (b) show the position of Berry curvature at K$^{\prime}$ point. The effect of biaxial strain on the Berry curvature in monolayer (c) H-BB WSiGeN$_4$ and (d) H-AA WSiGeP$_4$.}
	\label{p7}
\end{figure}

The transverse velocity can be amplified using the larger Berry curvature. Thus, faster charge transport and lesser recombination can be achieved. Therefore, a larger value of $\Omega_K$ is always desired, and Berry curvature modulation is indispensable in the field of valleytronics~\cite{tian2021manipulating}. Using Eqs.~\ref{e4} and~\ref{e5}, the Berry curvature of valence bands can be expressed as
\begin{equation}
	\Omega_v(\textbf{\textit{k}})=\tau\frac{2a^2t^2\Delta}{(4a^2t^2\textbf{\textit{k}}^2+\Delta^2)^{3/2}}
\end{equation}
where $a$, $t$ and $\Delta$ are the lattice constant, nearest-neighbor hopping integral, and direct band gap, respectively. Therefore, the Berry curvature can be enhanced by tuning the parameters $a$, $t$ and $\Delta$. Figure~\ref{p7}(a), \ref{p7}(b) and Table~\ref{T2} show the modulation of Berry curvature via different stacking orders. Surprisingly, the Berry curvature can be enhanced up to 400\% compared to its ground state. For the case of monolayer H-AA WSiGeP$_4$, it increases Berry curvature from 80 \AA$^2$ to 335 \AA$^2$ by changing the phase to the H-BB. Similarly, Berry curvature is also enhanced to 411 \AA$^2$ for monolayer H-BB WSiGeAs$_4$. However, modification is almost negligible in the case of monolayer H-BB WSiGeN$_4$ due to the small size of the N atom. The H-BB phase has the smallest value of $\Delta$ at the K point, thereby highest Berry curvature is obtained for this phase. The percentage change in the direct gap is highest in the case of the WSiGeAs$_4$ monolayer, therefore, the strongest tuning is obtained using symmetry control for WSiGeAs$_4$ monolayer. 

It is crucial to understand the underlying mechanism behind the Berry curvature modulation. The two main factors affecting SOC are macroscopic charge transfer and microscopic orbital contribution~\cite{zhou2021manipulation}. Table~\ref{T3} shows the orbital contribution at K point for the VB for monolayer WSiGeP$_4$. A similar trend is also observed for the monolayer WSiGeAs$_4$ and WSiGeN$_4$. The W-$d$ and P-$p$ orbitals contribute to valence states, where W-$d_{xy}$ and W-$d_{x^2-y^2}$ have the largest contributions. Rotation of Si/Ge-Z layers leads to a change in the orbital contribution, hence, the strength of the SOC. The contribution coming from heavy W-$d$ is highest in the case of the H-BB phase. Thus, the highest SOC strength leads to the largest strength of Berry curvature at the K/K$^{\prime}$ point for the H-BB phase. The contribution of W-$d$ orbitals follows the trend, H-BB $>$ H-BA $>$ H-AB $>$ H-AA and the same trend is followed by the Berry curvature. This shows that microscopic orbital contribution has a strong correlation with the Berry curvature. 

\begin{table*}[ht]
	\begin{center}
		\caption{Orbital contribution to the valence band at K/K$'$ point of monolayer H-AA WSiGeP$_4$.}
		\begin{tabular}{p{1.7cm} p{1.6cm} p{1.6cm} p{1.6cm} p{1.6cm} p{1.6cm} p{1.6cm} p{1.6cm} p{1.6cm}}
			\hline
			\hline
			Phase& $p_x$   &  $p_y$  & $p_z$ & $d_{xy}$ & $d_{yz}$ & $d_{z^2}$ & $d_{xz}$ & $d_{x^2-y^2}$   \\ \hline
			H-AA & 0.055 & 0.000 & 0.055 & 0.271 & 0.002 & 0.000  & 0.002 & 0.271  \\
			H-AB & 0.055 & 0.009 & 0.055 & 0.271 & 0.003 & 0.002  & 0.003 & 0.271 \\
			H-BA & 0.058 & 0.013 & 0.058 & 0.280 & 0.004 & 0.001  & 0.004 & 0.280 \\
			H-BB & 0.057 & 0.025 & 0.057 & 0.303 & 0.001 & 0.000  & 0.001 & 0.303 \\
			\hline\hline
		\end{tabular}
		\label{T3}
	\end{center}
\end{table*}

Another efficient way to modify the orbital contribution is by applying strain~\cite{tian2021manipulating,son2019strain}. Hence, the effect of tensile and compressive biaxial strain on the Berry curvature of VB is also analyzed. Figures~\ref{p7}(c) and \ref{p7}(d) show the Berry curvature of unstrained and strained cases for the VB in monolayer H-BB WSiGeN$_4$ and H-AA WSiGeP$_4$. The strain could also tune the Berry curvature efficiently. Under the tensile strain, the direct band gap and the magnitude of wavevector are reduced. Therefore, the magnitude of the Berry curvature increases. In the case of WSiGeN$_4$, Berry curvature increases from 15 \AA$^2$ to 25 \AA$^2$ under 5\% tensile strain. Whereas, it increases from 80 \AA$^2$ to 129 \AA$^2$ in WSiGeP$_4$ and is significantly smaller than the Berry curvature 335 \AA$^2$ obtained by symmetry control.

\subsection{Rashba effect}
\begin{figure}[ht]
	\includegraphics[width=8cm]{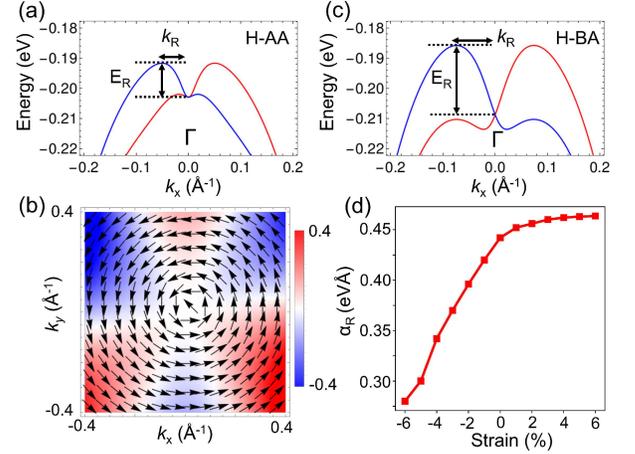}
	\caption{(a) The spin-projected band structures and (b) associated spin texture of top valence bands in H-AA WSiGeP$_4$. The arrows and color represent the in-plane ($x$, $y$) and out-of-plane ($z$) components of spin textures, respectively. (c) The spin-projected band structure of top valence bands in H-BA WSiGeP$_4$. (d) The variation of Rashba coefficients under the application of biaxial strain.}
	\label{p8}
\end{figure}

The presence of strong SOC in addition to the breaking of surface inversion symmetry along $z$-direction leads to the Rashba spin splitting of VB around $\Gamma$ point. The spin-projected band structure of top VBs of monolayer H-AA WSiGeP$_4$ is depicted in Fig.~\ref{p8}(a). The bands describe unique Rashba type dispersion, which can be explained by linear Rashba Hamiltonian. The helical nature of in-plane spin textures as shown in Fig.~\ref{p8}(b), confirms the presence of the Rashba effect around the $\Gamma$ point. Moreover, a small out-of-plane spin component has three-fold rotation symmetry in-line with the symmetry of the crystal.

\begin{figure*}[ht]
	\includegraphics[width=13cm]{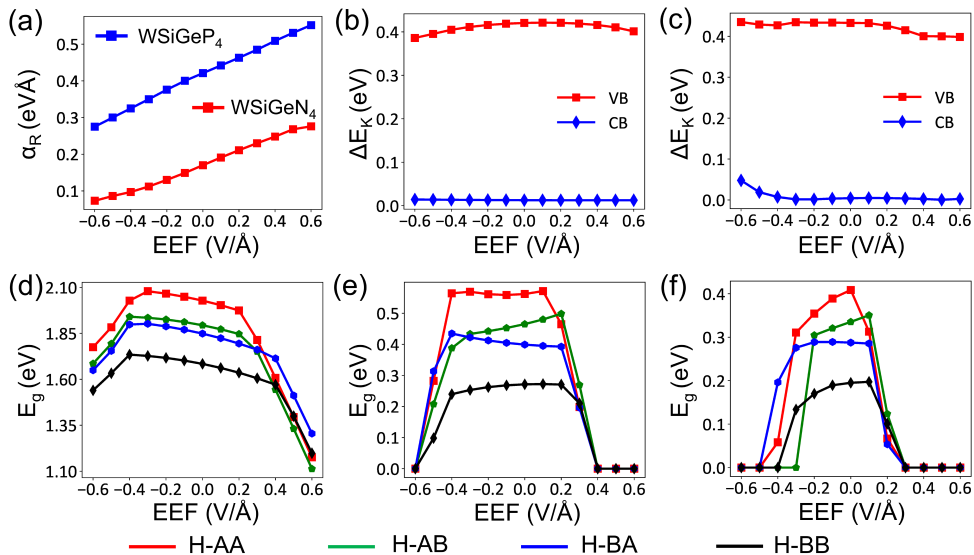}
	\caption{(a) The variation of Rashba coefficient ($\alpha_R$) under the application of out-of-plane EEF. Zeeman spin splittings at K/K$^{\prime}$ points as a function of EEF in monolayer (b) H-BB WSiGeN$_4$ and (c) H-AA WSiGeP$_4$. Modulation of band gaps with strain in monolayer (d) H-BB WSiGeN$_4$, (e) H-AA WSiGeP$_4$ and (f) H-AA WSiGeAs$_4$.}
	\label{p10}
\end{figure*}

To understand the band structure and spin texture, we have constructed a \textit{\textbf{k.p}} model Hamiltonian near the $\Gamma$ point. The little group of the $\Gamma$ point is $C_{3v}$, and the corresponding Hamiltonian is expressed as~\cite{yu2021spin,li2019intrinsic,bhumla2021origin}
\begin{equation}
	H_{\Gamma}(\textbf{\textit{k}})=H_o(\textbf{\textit{k}})+\alpha_R(k_x\sigma_y-k_y\sigma_x)+\gamma(k_y^3-3k_xk_y^2)
\end{equation}
where $H_o(k)$ is the free particle Hamiltonian, and $\alpha_R$ is the linear Rashba coefficient. $\gamma$ is the parameter signifying the contributions from the cubic term in the splitting and takes the out-of-plane spin contributions into consideration. We have observed that the energy contribution coming from the cubic term is much smaller than the linear term. The fitting of the DFT band structure by \textit{\textbf{k.p}} model hamiltonian allows the calculation of $\alpha_R$. Within the linear Rashba term, the $\alpha_R$ is also estimated by $\alpha_R=\frac{2E_R}{k_R}$, where $E_R$ and $k_R$ are the Rashba energy splitting and momentum shift, respectively~\cite{yu2021spin}. $E_R$ and $k_R$ are schematically shown in Fig.~\ref{p8}(a). The calculated values of $E_R$ and $k_R$ are $11.2$ meV and $0.55$ \AA$^{-1}$, respectively. Thus, for monolayer H-AA WSiGeP$_4$, $\alpha_R$ is 0.41 eV{\AA}. A larger value of $\alpha_R$ makes the Rashba effect more accessible experimentally. Therefore, greater magnitude of $\alpha_R$ is always desired. Consequently, we have studied the impact of symmetry on the strength of Rashba effect. Table~\ref{T4} shows $E_R$, $k_R$, and $\alpha_R$ of monolayer WSiGeN$_4$ and WSiGeP$_4$ in different configurations. H-BA structure has the largest value of $E_R$ and $\alpha_R$. As shown in Fig.~\ref{p8}(c), $E_R$ increases to $22.9$ meV from $11.2$ meV when the structure changes to H-BA from H-AA. The value of $\alpha_R$ is also significantly enhanced from 0.41 eV{\AA} to 0.61 eV{\AA}. Similar results have also been obtained for WSiGeN$_4$ and WSiGeAs$_4$. The observed trend of $\alpha_R$ is H-BA $>$ H-BB $>$ H-AA $>$ H-AB. 
                    
According to the Rashba model, spin splitting occurs in the plane perpendicular to the electric field. The net vertical electric field acting on the metal atom W, determines the strength of the Rashba effect. Therefore, the larger the out-of-plane intrinsic electric field, the more prominent is the in-plane spin splitting. In case of H-BA, the out-of-plane intrinsic electric field arising due to the formation of Janus and the position of Z atoms support each other. Therefore, the strongest Rashba spin splitting is observed for the H-BA phase as it has the highest out-of-plane asymmetry.

\begin{table}[ht]
	\begin{center}
		\caption{The calculated parameters ($E_R$, $k_R$ and $\alpha_R$) for Rashba spin splitting around $\Gamma$ point.}
		\begin{tabular}{p{1.8cm} p{1.3cm} p{1.6cm} p{1.6cm} p{1.6cm}}
			\hline
			\hline
			& Phase  & $ E_R$ (meV)  & $k_R$ (\AA$^{-1}$) & $\alpha_R$ (eV\AA) \\ \hline  
			WSiGeN$_4$ & H-AA  & 5.7 & 0.060 &  0.19 \\
			& H-AB  & 3.0 & 0.068 &  0.09    \\
			& H-BA  & 9.4 & 0.067 & 0.28   \\
			& H-BB  & 5.6 & 0.066 & 0.17    \\
			WSiGeP$_4$ & H-AA  & 11.2  & 0.055  & 0.41  \\
			& H-AB  & 8.57 & 0.079 & 0.22  \\
			& H-BA  & 22.9  & 0.075 & 0.61  \\
			& H-BB  & 16.9 & 0.079 &  0.43  \\
			\hline\hline
		\end{tabular}
		\label{T4}
	\end{center}
	
\end{table}
\begin{table*}[ht]
	\begin{center}
		\caption{The acoustic phonon limited carrier mobility and other relevant parameters for WSiGeZ$_4$ monolayer. The $x$ and $y$ directions correspond to armchair and zigzag directions, respectively. $m_o$ is rest mass of the electron.}
		\begin{tabular}{p{2cm} p{1.4cm} p{1.5cm} p{1.5cm} p{1.5cm}  p{1.5cm} p{1.5cm} p{2cm} p{2cm} }
			\hline
			\hline
			& Carrier  & $m_x$/$m_o$  & $m_y$/$m_o$ & $E_x$ & $E_y$& $C_{2D}$ & $\mu_x$ & $\mu_y$\\
			&  &   &  & (eV) & (eV)& (Nm$^{-1}$) & (cm$^{2}$V$^{-1}$s$^{-1}$) & (cm$^{2}$V$^{-1}$s$^{-1}$) \\ \hline  
			WSiGeN$_4$ & $e_{K}$       & 0.460   & 0.499  &  5.82  & 6.832 & 493.1 & 2355.1 & 942.2 \\
			& $h_{\Gamma}$  & 1.838   & 1.816  &  5.70  & 5.147 & 493.1 & 98.3 & 119.7 \\
			WSiGeP$_4$ & $e_{K}$       & 0.975   & 0.594  &  13.18 & 10.32 & 225.5 & 37.3 & 99.9 \\
			& $h_{K}$       & 0.484   & 0.448  &   9.49 & 6.98  & 225.5 & 237.0 & 473.5\\
			WSiGeAs$_4$& $e_{K}$       & 3.260   & 1.673  &  11.92 & 10.88 & 178.2 & 3.5 & 8.2 \\
			& $h_{K}$       & 0.542   & 0.477  &   9.59 & 8.91  & 178.2 & 150.1 &  197.6\\
			\hline\hline
		\end{tabular}
		\label{T5}
	\end{center}
\end{table*}
Additionally, the effect of biaxial strain on $\alpha_R$ is also studied, and the variation is shown in Fig.~\ref{p8}(d). The value of $\alpha_R$ decreases from 0.41 eV\AA\ to 0.28 eV\AA\ under a compressive strain of 6\%. The tensile strain of 6\% increase $\alpha_R$ to 0.46 eV\textrm{$\AA$}, which is significantly smaller than the increment observed using the symmetry control. The strain is also an effective way to modulate the band edge positions and band gaps. The complete band structures under the application of biaxial strain are shown in section VI of SM. For the case of WSiGeN$_4$, the band gap remains indirect in the range of -6\% to 6\%. The band gap remains direct for WSiGeP$_4$ and WSiGeAs$_4$ in the range of -3\% to 3\%. In general, the band gap shows decrement under tensile strain. However, under compressive strain, the band gap increases up to 2\% and decreases afterward. Furthermore, WSiGeAs$_4$ shows semiconductor to metal transitions under a compressive strain of larger than 3\%.
\subsection{The role of electric field}
An external electric field (EEF) is commonly used to modify the strength of Rashba effect in spintronics devices~\cite{cheng2016nonlinear,yao2017manipulation}. Therefore, the effect of EEF is also investigated in this study. We have applied the out-of-plane EEF in the range of -0.6 V/\AA\, to 0.6 V/\AA\,to analyze the variation of Rashba coefficient ($\alpha_R$) and Zeeman splittings ($\Delta E_{K}^{CB}$ and $\Delta E_{K}^{VB}$). Figure~\ref{p10}(a) shows the variation of $\alpha_R$ as a function of EEF for monolayer H-BB WSiGeN$_4$ and H-AA WSiGeP$_4$. When the EEF points from Ge to Si (taken to be positive direction), which is consistent with the local electric field, it enhances the Rashba SOC. Whereas, the opposite direction of EEF reduces the strength of Rashba SOC, as the effects of the EEF and the local electric field cancel each other. The value of $\alpha_R$ varies from 0.28 eV{\AA} to 0.55 eV/{\AA} in case of monolayer H-AA WSiGeP$_4$, while it varies from 0.08 eV/{\AA} to 0.25 eV/{\AA} for monolayer H-BB WSiGeN$_4$ . The variation of Zeeman splittings at the K point are shown in Figs.~\ref{p10}(b) and \ref{p10}(c). The $d_{x^2-y^2}$ and $d_{xy}$ orbitals under the influence of in-plane electric field contribute to the Zeeman splitting at K point. Therefore, the out-of-plane EEF hardly affects the $\Delta E_{K}^{CB}$ and $\Delta E_{K}^{VB}$. The evolution of band gap as a function of EEF is provided in Figs.~\ref{p10}(d), ~\ref{p10}(e) and ~\ref{p10}(f) (see section V of SM for band structures). We have noticed that the band gaps are almost insensitive to small EEFs (in the range -0.3 V/\AA\, to 0.3 V/\AA). It is also interesting to note that the CBM for monolayer WSiGeZ$_4$ shifts to the M point under large EEF (see section VI of SM), this is because the Si atom starts contributing to the CBM. Therefore, band gaps decrease sharply under large EEFs. In the case of WSiGeP$_4$ and WSiGeAs$_4$, giant Stark effect closes the band gap and semiconductor to metal transition takes place.

\subsection{Carrier mobility}
Estimation of carrier mobilities is essential for exploring the potential use of any material for application in electronic devices~\cite{priydarshi2022large,guo2021predicted}. The carrier mobility ($\mu_{2D}$) of the WSiGeZ$_4$ monolayer is calculated using the deformation potential (DP) theory. According to DP theory~\cite{bruzzone2011ab}, $\mu_{2D}$ is expressed as
\begin{equation}
	\mu_{2D}=\frac{e\hbar^3C_{2D}}{k_BTm^*m_dE_l^2 }
\end{equation}
where $m_d$ and $m^*$ are the average effective mass ($\sqrt{m_xm_y}$) and the effective mass along the transport direction, respectively. $C_{2D}$ is Young's elastic modulus, which is calculated from the elastic tensor $C_{ij}$. The $E_l$ is the DP constant defined by $E_l=\delta E/\delta l$. Here, $\delta E$ is the change in the position of the CBM or VBM with respect to the vacuum level and $\delta l=\Delta l/l$ is the relative strain. The detailed analysis of the method used to compute elastic constants and carrier mobilities is discussed in section VI of SM. The calculated elastic constants and carrier mobilities are listed in Table~\ref{T5}. The carrier mobilities in these monolayers are anisotropic and show considerable variation. For instance, the most significant carrier mobility observed is 2355.1 cm$^2$V$^{-1}$s$^{-1}$ for the electron in monolayer H-BB WSiGeN$_4$, and the smallest carrier mobility observed is as tiny as 3.514 cm$^2$V$^{-1}$s$^{-1}$ for the electron in monolayer H-AA WSiGeAs$_4$.

\section{Conclusion}
 We have revealed that different stacking patterns of Si-Z, Z-W-Z and Ge-Z generate the dynamically stable structures. The most stable structure for monolayer WSiGeN$_4$ is H-BB, whereas for monolayer WSiGeP$_4$ and WSiGeAs$_4$ is H-AA. These materials show valley contrasting circular dichroism, spin-valley coupling, and Berry curvature. Additionally, the out-of-plane asymmetry in them leads to Rashba effect around the center of the BZ. The stacking pattern greatly modifies the spintronics and valleytronics properties. The highest Rashba spin-orbit coupling around the $\Gamma$ point is observed for the H-BA phase having the highest out-of-plane electric dipole moment. Whereas, the Berry curvature is strongest at valleys for H-BB phase with the highest contribution coming from $d_{xy}$ and $d_{x^2-y^2}$ orbitals. Additionally, valleys observed in WSiGeP$_4$ and WSiGeAs$_4$ monolayers at K and K$'$ points have multiple folds.  Furthermore, we have found that the band gap type and valleys can be tuned with an EEF and biaxial strain. Under large strain and high electric field, monolayer WSiGeP$_4$ and WSiGeAs$_4$ show semiconductor to metal transition. We have found the strong anisotropy in carrier mobilities along the armchair and zigzag directions. Rashba spin splitting and valley polarization can facilitate the manipulation of electron spin and valley. The regulation of spin direction of separated valley carriers by controlling symmetry can be realized better in septuple-atomic-layer materials compared to TMDs.

\section*{\MakeUppercase{Acknowledgements}}
S.S. acknowledges CSIR, India, for the senior research fellowship [grant no. 09/086(1432)/2019-EMR-I]. A.P. acknowledges IIT Delhi for the senior research fellowship. R.M. acknowledges CSIR, India, for the junior research fellowship [grant no. 09/0086(12865)/2021-EMR-I]. S. B. acknowledges financial support from SERB under a core research grant (grant no. CRG/2019/000647) to set up his High Performance Computing (HPC) facility ``Veena" at IIT Delhi for computational resources.
\bibliography{references}

\begin{thebibliography}{67}%
\makeatletter
\providecommand \@ifxundefined [1]{%
 \@ifx{#1\undefined}
}%
\providecommand \@ifnum [1]{%
 \ifnum #1\expandafter \@firstoftwo
 \else \expandafter \@secondoftwo
 \fi
}%
\providecommand \@ifx [1]{%
 \ifx #1\expandafter \@firstoftwo
 \else \expandafter \@secondoftwo
 \fi
}%
\providecommand \natexlab [1]{#1}%
\providecommand \enquote  [1]{``#1''}%
\providecommand \bibnamefont  [1]{#1}%
\providecommand \bibfnamefont [1]{#1}%
\providecommand \citenamefont [1]{#1}%
\providecommand \href@noop [0]{\@secondoftwo}%
\providecommand \href [0]{\begingroup \@sanitize@url \@href}%
\providecommand \@href[1]{\@@startlink{#1}\@@href}%
\providecommand \@@href[1]{\endgroup#1\@@endlink}%
\providecommand \@sanitize@url [0]{\catcode `\\12\catcode `\$12\catcode
  `\&12\catcode `\#12\catcode `\^12\catcode `\_12\catcode `\%12\relax}%
\providecommand \@@startlink[1]{}%
\providecommand \@@endlink[0]{}%
\providecommand \url  [0]{\begingroup\@sanitize@url \@url }%
\providecommand \@url [1]{\endgroup\@href {#1}{\urlprefix }}%
\providecommand \urlprefix  [0]{URL }%
\providecommand \Eprint [0]{\href }%
\providecommand \doibase [0]{http://dx.doi.org/}%
\providecommand \selectlanguage [0]{\@gobble}%
\providecommand \bibinfo  [0]{\@secondoftwo}%
\providecommand \bibfield  [0]{\@secondoftwo}%
\providecommand \translation [1]{[#1]}%
\providecommand \BibitemOpen [0]{}%
\providecommand \bibitemStop [0]{}%
\providecommand \bibitemNoStop [0]{.\EOS\space}%
\providecommand \EOS [0]{\spacefactor3000\relax}%
\providecommand \BibitemShut  [1]{\csname bibitem#1\endcsname}%
\let\auto@bib@innerbib\@empty
\bibitem [{\citenamefont {Geim}(2009)}]{geim2009graphene}%
  \BibitemOpen
  \bibfield  {author} {\bibinfo {author} {\bibfnamefont {A.~K.}\ \bibnamefont
  {Geim}},\ }\href@noop {} {\bibfield  {journal} {\bibinfo  {journal}
  {Science}\ }\textbf {\bibinfo {volume} {324}},\ \bibinfo {pages} {1530}
  (\bibinfo {year} {2009})}\BibitemShut {NoStop}%
\bibitem [{\citenamefont {Novoselov}\ \emph {et~al.}(2012)\citenamefont
  {Novoselov}, \citenamefont {Colombo}, \citenamefont {Gellert}, \citenamefont
  {Schwab},\ and\ \citenamefont {Kim}}]{novoselov2012roadmap}%
  \BibitemOpen
  \bibfield  {author} {\bibinfo {author} {\bibfnamefont {K.~S.}\ \bibnamefont
  {Novoselov}}, \bibinfo {author} {\bibfnamefont {L.}~\bibnamefont {Colombo}},
  \bibinfo {author} {\bibfnamefont {P.}~\bibnamefont {Gellert}}, \bibinfo
  {author} {\bibfnamefont {M.}~\bibnamefont {Schwab}}, \ and\ \bibinfo {author}
  {\bibfnamefont {K.}~\bibnamefont {Kim}},\ }\href@noop {} {\bibfield
  {journal} {\bibinfo  {journal} {Nature}\ }\textbf {\bibinfo {volume} {490}},\
  \bibinfo {pages} {192} (\bibinfo {year} {2012})}\BibitemShut {NoStop}%
\bibitem [{\citenamefont {Kane}\ and\ \citenamefont
  {Mele}(2005)}]{kane2005quantum}%
  \BibitemOpen
  \bibfield  {author} {\bibinfo {author} {\bibfnamefont {C.~L.}\ \bibnamefont
  {Kane}}\ and\ \bibinfo {author} {\bibfnamefont {E.~J.}\ \bibnamefont
  {Mele}},\ }\href@noop {} {\bibfield  {journal} {\bibinfo  {journal} {Phys.
  Rev. Lett.}\ }\textbf {\bibinfo {volume} {95}},\ \bibinfo {pages} {226801}
  (\bibinfo {year} {2005})}\BibitemShut {NoStop}%
\bibitem [{\citenamefont {Molle}\ \emph {et~al.}(2018)\citenamefont {Molle},
  \citenamefont {Grazianetti}, \citenamefont {Tao}, \citenamefont {Taneja},
  \citenamefont {Alam},\ and\ \citenamefont {Akinwande}}]{molle2018silicene}%
  \BibitemOpen
  \bibfield  {author} {\bibinfo {author} {\bibfnamefont {A.}~\bibnamefont
  {Molle}}, \bibinfo {author} {\bibfnamefont {C.}~\bibnamefont {Grazianetti}},
  \bibinfo {author} {\bibfnamefont {L.}~\bibnamefont {Tao}}, \bibinfo {author}
  {\bibfnamefont {D.}~\bibnamefont {Taneja}}, \bibinfo {author} {\bibfnamefont
  {M.~H.}\ \bibnamefont {Alam}}, \ and\ \bibinfo {author} {\bibfnamefont
  {D.}~\bibnamefont {Akinwande}},\ }\href@noop {} {\bibfield  {journal}
  {\bibinfo  {journal} {Chem. Soc. Rev.}\ }\textbf {\bibinfo {volume} {47}},\
  \bibinfo {pages} {6370} (\bibinfo {year} {2018})}\BibitemShut {NoStop}%
\bibitem [{\citenamefont {Ni}\ \emph {et~al.}(2012)\citenamefont {Ni},
  \citenamefont {Liu}, \citenamefont {Tang}, \citenamefont {Zheng},
  \citenamefont {Zhou}, \citenamefont {Qin}, \citenamefont {Gao}, \citenamefont
  {Yu},\ and\ \citenamefont {Lu}}]{ni2012tunable}%
  \BibitemOpen
  \bibfield  {author} {\bibinfo {author} {\bibfnamefont {Z.}~\bibnamefont
  {Ni}}, \bibinfo {author} {\bibfnamefont {Q.}~\bibnamefont {Liu}}, \bibinfo
  {author} {\bibfnamefont {K.}~\bibnamefont {Tang}}, \bibinfo {author}
  {\bibfnamefont {J.}~\bibnamefont {Zheng}}, \bibinfo {author} {\bibfnamefont
  {J.}~\bibnamefont {Zhou}}, \bibinfo {author} {\bibfnamefont {R.}~\bibnamefont
  {Qin}}, \bibinfo {author} {\bibfnamefont {Z.}~\bibnamefont {Gao}}, \bibinfo
  {author} {\bibfnamefont {D.}~\bibnamefont {Yu}}, \ and\ \bibinfo {author}
  {\bibfnamefont {J.}~\bibnamefont {Lu}},\ }\href@noop {} {\bibfield  {journal}
  {\bibinfo  {journal} {Nano Lett.}\ }\textbf {\bibinfo {volume} {12}},\
  \bibinfo {pages} {113} (\bibinfo {year} {2012})}\BibitemShut {NoStop}%
\bibitem [{\citenamefont {Chopra}\ \emph {et~al.}(1995)\citenamefont {Chopra},
  \citenamefont {Luyken}, \citenamefont {Cherrey}, \citenamefont {Crespi},
  \citenamefont {Cohen}, \citenamefont {Louie},\ and\ \citenamefont
  {Zettl}}]{chopra1995boron}%
  \BibitemOpen
  \bibfield  {author} {\bibinfo {author} {\bibfnamefont {N.~G.}\ \bibnamefont
  {Chopra}}, \bibinfo {author} {\bibfnamefont {R.}~\bibnamefont {Luyken}},
  \bibinfo {author} {\bibfnamefont {K.}~\bibnamefont {Cherrey}}, \bibinfo
  {author} {\bibfnamefont {V.~H.}\ \bibnamefont {Crespi}}, \bibinfo {author}
  {\bibfnamefont {M.~L.}\ \bibnamefont {Cohen}}, \bibinfo {author}
  {\bibfnamefont {S.~G.}\ \bibnamefont {Louie}}, \ and\ \bibinfo {author}
  {\bibfnamefont {A.}~\bibnamefont {Zettl}},\ }\href@noop {} {\bibfield
  {journal} {\bibinfo  {journal} {Science}\ }\textbf {\bibinfo {volume}
  {269}},\ \bibinfo {pages} {966} (\bibinfo {year} {1995})}\BibitemShut
  {NoStop}%
\bibitem [{\citenamefont {Manzeli}\ \emph {et~al.}(2017)\citenamefont
  {Manzeli}, \citenamefont {Ovchinnikov}, \citenamefont {Pasquier},
  \citenamefont {Yazyev},\ and\ \citenamefont {Kis}}]{manzeli20172d}%
  \BibitemOpen
  \bibfield  {author} {\bibinfo {author} {\bibfnamefont {S.}~\bibnamefont
  {Manzeli}}, \bibinfo {author} {\bibfnamefont {D.}~\bibnamefont
  {Ovchinnikov}}, \bibinfo {author} {\bibfnamefont {D.}~\bibnamefont
  {Pasquier}}, \bibinfo {author} {\bibfnamefont {O.~V.}\ \bibnamefont
  {Yazyev}}, \ and\ \bibinfo {author} {\bibfnamefont {A.}~\bibnamefont {Kis}},\
  }\href@noop {} {\bibfield  {journal} {\bibinfo  {journal} {Nat. Rev. Mater.}\
  }\textbf {\bibinfo {volume} {2}},\ \bibinfo {pages} {1} (\bibinfo {year}
  {2017})}\BibitemShut {NoStop}%
\bibitem [{\citenamefont {Xu}\ \emph {et~al.}(2014)\citenamefont {Xu},
  \citenamefont {Yao}, \citenamefont {Xiao},\ and\ \citenamefont
  {Heinz}}]{xu2014spin}%
  \BibitemOpen
  \bibfield  {author} {\bibinfo {author} {\bibfnamefont {X.}~\bibnamefont
  {Xu}}, \bibinfo {author} {\bibfnamefont {W.}~\bibnamefont {Yao}}, \bibinfo
  {author} {\bibfnamefont {D.}~\bibnamefont {Xiao}}, \ and\ \bibinfo {author}
  {\bibfnamefont {T.~F.}\ \bibnamefont {Heinz}},\ }\href@noop {} {\bibfield
  {journal} {\bibinfo  {journal} {Nat. Phys.}\ }\textbf {\bibinfo {volume}
  {10}},\ \bibinfo {pages} {343} (\bibinfo {year} {2014})}\BibitemShut
  {NoStop}%
\bibitem [{\citenamefont {Wang}\ \emph {et~al.}(2018)\citenamefont {Wang},
  \citenamefont {Chernikov}, \citenamefont {Glazov}, \citenamefont {Heinz},
  \citenamefont {Marie}, \citenamefont {Amand},\ and\ \citenamefont
  {Urbaszek}}]{wang2018colloquium}%
  \BibitemOpen
  \bibfield  {author} {\bibinfo {author} {\bibfnamefont {G.}~\bibnamefont
  {Wang}}, \bibinfo {author} {\bibfnamefont {A.}~\bibnamefont {Chernikov}},
  \bibinfo {author} {\bibfnamefont {M.~M.}\ \bibnamefont {Glazov}}, \bibinfo
  {author} {\bibfnamefont {T.~F.}\ \bibnamefont {Heinz}}, \bibinfo {author}
  {\bibfnamefont {X.}~\bibnamefont {Marie}}, \bibinfo {author} {\bibfnamefont
  {T.}~\bibnamefont {Amand}}, \ and\ \bibinfo {author} {\bibfnamefont
  {B.}~\bibnamefont {Urbaszek}},\ }\href@noop {} {\bibfield  {journal}
  {\bibinfo  {journal} {Rev. Mod. Phys.}\ }\textbf {\bibinfo {volume} {90}},\
  \bibinfo {pages} {021001} (\bibinfo {year} {2018})}\BibitemShut {NoStop}%
\bibitem [{\citenamefont {Liang}\ \emph {et~al.}(2014)\citenamefont {Liang},
  \citenamefont {Zhang},\ and\ \citenamefont {Yang}}]{liang2014rational}%
  \BibitemOpen
  \bibfield  {author} {\bibinfo {author} {\bibfnamefont {F.}~\bibnamefont
  {Liang}}, \bibinfo {author} {\bibfnamefont {C.}~\bibnamefont {Zhang}}, \ and\
  \bibinfo {author} {\bibfnamefont {Z.}~\bibnamefont {Yang}},\ }\href@noop {}
  {\bibfield  {journal} {\bibinfo  {journal} {Adv. Mater.}\ }\textbf {\bibinfo
  {volume} {26}},\ \bibinfo {pages} {6944} (\bibinfo {year}
  {2014})}\BibitemShut {NoStop}%
\bibitem [{\citenamefont {Li}\ \emph {et~al.}(2018)\citenamefont {Li},
  \citenamefont {Cheng},\ and\ \citenamefont {Huang}}]{li2018recent}%
  \BibitemOpen
  \bibfield  {author} {\bibinfo {author} {\bibfnamefont {R.}~\bibnamefont
  {Li}}, \bibinfo {author} {\bibfnamefont {Y.}~\bibnamefont {Cheng}}, \ and\
  \bibinfo {author} {\bibfnamefont {W.}~\bibnamefont {Huang}},\ }\href@noop {}
  {\bibfield  {journal} {\bibinfo  {journal} {Small}\ }\textbf {\bibinfo
  {volume} {14}},\ \bibinfo {pages} {1802091} (\bibinfo {year}
  {2018})}\BibitemShut {NoStop}%
\bibitem [{\citenamefont {Hu}\ \emph {et~al.}(2018)\citenamefont {Hu},
  \citenamefont {Jia}, \citenamefont {Zhao}, \citenamefont {Wu}, \citenamefont
  {Stroppa},\ and\ \citenamefont {Ren}}]{hu2018intrinsic}%
  \BibitemOpen
  \bibfield  {author} {\bibinfo {author} {\bibfnamefont {T.}~\bibnamefont
  {Hu}}, \bibinfo {author} {\bibfnamefont {F.}~\bibnamefont {Jia}}, \bibinfo
  {author} {\bibfnamefont {G.}~\bibnamefont {Zhao}}, \bibinfo {author}
  {\bibfnamefont {J.}~\bibnamefont {Wu}}, \bibinfo {author} {\bibfnamefont
  {A.}~\bibnamefont {Stroppa}}, \ and\ \bibinfo {author} {\bibfnamefont
  {W.}~\bibnamefont {Ren}},\ }\href@noop {} {\bibfield  {journal} {\bibinfo
  {journal} {Phys. Rev. B}\ }\textbf {\bibinfo {volume} {97}},\ \bibinfo
  {pages} {235404} (\bibinfo {year} {2018})}\BibitemShut {NoStop}%
\bibitem [{\citenamefont {Touski}\ and\ \citenamefont
  {Ghobadi}(2021)}]{touski2021structural}%
  \BibitemOpen
  \bibfield  {author} {\bibinfo {author} {\bibfnamefont {S.~B.}\ \bibnamefont
  {Touski}}\ and\ \bibinfo {author} {\bibfnamefont {N.}~\bibnamefont
  {Ghobadi}},\ }\href@noop {} {\bibfield  {journal} {\bibinfo  {journal} {Phys.
  Rev. B}\ }\textbf {\bibinfo {volume} {103}},\ \bibinfo {pages} {165404}
  (\bibinfo {year} {2021})}\BibitemShut {NoStop}%
\bibitem [{\citenamefont {Singh}\ \emph {et~al.}(2021)\citenamefont {Singh},
  \citenamefont {Jain},\ and\ \citenamefont {Bhattacharya}}]{singh2021mos}%
  \BibitemOpen
  \bibfield  {author} {\bibinfo {author} {\bibfnamefont {A.}~\bibnamefont
  {Singh}}, \bibinfo {author} {\bibfnamefont {M.}~\bibnamefont {Jain}}, \ and\
  \bibinfo {author} {\bibfnamefont {S.}~\bibnamefont {Bhattacharya}},\
  }\href@noop {} {\bibfield  {journal} {\bibinfo  {journal} {Nanoscale Adv.}\
  }\textbf {\bibinfo {volume} {3}},\ \bibinfo {pages} {2837} (\bibinfo {year}
  {2021})}\BibitemShut {NoStop}%
\bibitem [{\citenamefont {Yuan}\ \emph {et~al.}(2020)\citenamefont {Yuan},
  \citenamefont {Yang}, \citenamefont {Cai}, \citenamefont {Wu}, \citenamefont
  {Chen}, \citenamefont {Yan}, \citenamefont {Shen} \emph
  {et~al.}}]{yuan2020intrinsic}%
  \BibitemOpen
  \bibfield  {author} {\bibinfo {author} {\bibfnamefont {J.}~\bibnamefont
  {Yuan}}, \bibinfo {author} {\bibfnamefont {Y.}~\bibnamefont {Yang}}, \bibinfo
  {author} {\bibfnamefont {Y.}~\bibnamefont {Cai}}, \bibinfo {author}
  {\bibfnamefont {Y.}~\bibnamefont {Wu}}, \bibinfo {author} {\bibfnamefont
  {Y.}~\bibnamefont {Chen}}, \bibinfo {author} {\bibfnamefont {X.}~\bibnamefont
  {Yan}}, \bibinfo {author} {\bibfnamefont {L.}~\bibnamefont {Shen}},  \emph
  {et~al.},\ }\href@noop {} {\bibfield  {journal} {\bibinfo  {journal}
  {Physical Review B}\ }\textbf {\bibinfo {volume} {101}},\ \bibinfo {pages}
  {094420} (\bibinfo {year} {2020})}\BibitemShut {NoStop}%
\bibitem [{\citenamefont {Dong}\ \emph {et~al.}(2017)\citenamefont {Dong},
  \citenamefont {Lou},\ and\ \citenamefont {Shenoy}}]{dong2017large}%
  \BibitemOpen
  \bibfield  {author} {\bibinfo {author} {\bibfnamefont {L.}~\bibnamefont
  {Dong}}, \bibinfo {author} {\bibfnamefont {J.}~\bibnamefont {Lou}}, \ and\
  \bibinfo {author} {\bibfnamefont {V.~B.}\ \bibnamefont {Shenoy}},\
  }\href@noop {} {\bibfield  {journal} {\bibinfo  {journal} {ACS nano}\
  }\textbf {\bibinfo {volume} {11}},\ \bibinfo {pages} {8242} (\bibinfo {year}
  {2017})}\BibitemShut {NoStop}%
\bibitem [{\citenamefont {Lu}\ \emph {et~al.}(2017)\citenamefont {Lu},
  \citenamefont {Zhu}, \citenamefont {Xiao}, \citenamefont {Chuu},
  \citenamefont {Han}, \citenamefont {Chiu}, \citenamefont {Cheng},
  \citenamefont {Yang}, \citenamefont {Wei}, \citenamefont {Yang} \emph
  {et~al.}}]{lu2017janus}%
  \BibitemOpen
  \bibfield  {author} {\bibinfo {author} {\bibfnamefont {A.-Y.}\ \bibnamefont
  {Lu}}, \bibinfo {author} {\bibfnamefont {H.}~\bibnamefont {Zhu}}, \bibinfo
  {author} {\bibfnamefont {J.}~\bibnamefont {Xiao}}, \bibinfo {author}
  {\bibfnamefont {C.-P.}\ \bibnamefont {Chuu}}, \bibinfo {author}
  {\bibfnamefont {Y.}~\bibnamefont {Han}}, \bibinfo {author} {\bibfnamefont
  {M.-H.}\ \bibnamefont {Chiu}}, \bibinfo {author} {\bibfnamefont {C.-C.}\
  \bibnamefont {Cheng}}, \bibinfo {author} {\bibfnamefont {C.-W.}\ \bibnamefont
  {Yang}}, \bibinfo {author} {\bibfnamefont {K.-H.}\ \bibnamefont {Wei}},
  \bibinfo {author} {\bibfnamefont {Y.}~\bibnamefont {Yang}},  \emph {et~al.},\
  }\href@noop {} {\bibfield  {journal} {\bibinfo  {journal} {Nature
  nanotechnology}\ }\textbf {\bibinfo {volume} {12}},\ \bibinfo {pages} {744}
  (\bibinfo {year} {2017})}\BibitemShut {NoStop}%
\bibitem [{\citenamefont {Jang}\ \emph {et~al.}(2022)\citenamefont {Jang},
  \citenamefont {Lee}, \citenamefont {Kim}, \citenamefont {Park}, \citenamefont
  {Kim},\ and\ \citenamefont {Choi}}]{jang2022growth}%
  \BibitemOpen
  \bibfield  {author} {\bibinfo {author} {\bibfnamefont {C.~W.}\ \bibnamefont
  {Jang}}, \bibinfo {author} {\bibfnamefont {W.~J.}\ \bibnamefont {Lee}},
  \bibinfo {author} {\bibfnamefont {J.~K.}\ \bibnamefont {Kim}}, \bibinfo
  {author} {\bibfnamefont {S.~M.}\ \bibnamefont {Park}}, \bibinfo {author}
  {\bibfnamefont {S.}~\bibnamefont {Kim}}, \ and\ \bibinfo {author}
  {\bibfnamefont {S.-H.}\ \bibnamefont {Choi}},\ }\href@noop {} {\bibfield
  {journal} {\bibinfo  {journal} {NPG Asia Materials}\ }\textbf {\bibinfo
  {volume} {14}},\ \bibinfo {pages} {1} (\bibinfo {year} {2022})}\BibitemShut
  {NoStop}%
\bibitem [{\citenamefont {Wang}\ \emph {et~al.}(2021)\citenamefont {Wang},
  \citenamefont {Shi}, \citenamefont {Liu}, \citenamefont {Zhang},
  \citenamefont {Hong}, \citenamefont {Li}, \citenamefont {Gao}, \citenamefont
  {Chen}, \citenamefont {Ren},\ and\ \citenamefont
  {Cheng}}]{wang2021intercalated}%
  \BibitemOpen
  \bibfield  {author} {\bibinfo {author} {\bibfnamefont {L.}~\bibnamefont
  {Wang}}, \bibinfo {author} {\bibfnamefont {Y.}~\bibnamefont {Shi}}, \bibinfo
  {author} {\bibfnamefont {M.}~\bibnamefont {Liu}}, \bibinfo {author}
  {\bibfnamefont {A.}~\bibnamefont {Zhang}}, \bibinfo {author} {\bibfnamefont
  {Y.-L.}\ \bibnamefont {Hong}}, \bibinfo {author} {\bibfnamefont
  {R.}~\bibnamefont {Li}}, \bibinfo {author} {\bibfnamefont {Q.}~\bibnamefont
  {Gao}}, \bibinfo {author} {\bibfnamefont {M.}~\bibnamefont {Chen}}, \bibinfo
  {author} {\bibfnamefont {W.}~\bibnamefont {Ren}}, \ and\ \bibinfo {author}
  {\bibfnamefont {H.-M.}\ \bibnamefont {Cheng}},\ }\href@noop {} {\bibfield
  {journal} {\bibinfo  {journal} {Nat. Commun.}\ }\textbf {\bibinfo {volume}
  {12}},\ \bibinfo {pages} {1} (\bibinfo {year} {2021})}\BibitemShut {NoStop}%
\bibitem [{\citenamefont {Hong}\ \emph {et~al.}(2020)\citenamefont {Hong},
  \citenamefont {Liu}, \citenamefont {Wang}, \citenamefont {Zhou},
  \citenamefont {Ma}, \citenamefont {Xu}, \citenamefont {Feng}, \citenamefont
  {Chen}, \citenamefont {Chen},\ and\ \citenamefont {Sun}}]{hong2020chemical}%
  \BibitemOpen
  \bibfield  {author} {\bibinfo {author} {\bibfnamefont {Y.-L.}\ \bibnamefont
  {Hong}}, \bibinfo {author} {\bibfnamefont {Z.}~\bibnamefont {Liu}}, \bibinfo
  {author} {\bibfnamefont {L.}~\bibnamefont {Wang}}, \bibinfo {author}
  {\bibfnamefont {T.}~\bibnamefont {Zhou}}, \bibinfo {author} {\bibfnamefont
  {W.}~\bibnamefont {Ma}}, \bibinfo {author} {\bibfnamefont {C.}~\bibnamefont
  {Xu}}, \bibinfo {author} {\bibfnamefont {S.}~\bibnamefont {Feng}}, \bibinfo
  {author} {\bibfnamefont {L.}~\bibnamefont {Chen}}, \bibinfo {author}
  {\bibfnamefont {M.-L.}\ \bibnamefont {Chen}}, \ and\ \bibinfo {author}
  {\bibfnamefont {D.-M.}\ \bibnamefont {Sun}},\ }\href@noop {} {\bibfield
  {journal} {\bibinfo  {journal} {Science}\ }\textbf {\bibinfo {volume}
  {369}},\ \bibinfo {pages} {670} (\bibinfo {year} {2020})}\BibitemShut
  {NoStop}%
\bibitem [{\citenamefont {Guo}\ \emph {et~al.}(2021{\natexlab{a}})\citenamefont
  {Guo}, \citenamefont {Zhu}, \citenamefont {Mu},\ and\ \citenamefont
  {Chen}}]{guo2021piezoelectric}%
  \BibitemOpen
  \bibfield  {author} {\bibinfo {author} {\bibfnamefont {S.-D.}\ \bibnamefont
  {Guo}}, \bibinfo {author} {\bibfnamefont {Y.-T.}\ \bibnamefont {Zhu}},
  \bibinfo {author} {\bibfnamefont {W.-Q.}\ \bibnamefont {Mu}}, \ and\ \bibinfo
  {author} {\bibfnamefont {X.-Q.}\ \bibnamefont {Chen}},\ }\href@noop {}
  {\bibfield  {journal} {\bibinfo  {journal} {Journal of Materials Chemistry
  C}\ }\textbf {\bibinfo {volume} {9}},\ \bibinfo {pages} {7465} (\bibinfo
  {year} {2021}{\natexlab{a}})}\BibitemShut {NoStop}%
\bibitem [{\citenamefont {Rezavand}\ \emph {et~al.}(2022)\citenamefont
  {Rezavand}, \citenamefont {Ghobadi},\ and\ \citenamefont
  {Behnamghader}}]{rezavand2022electronic}%
  \BibitemOpen
  \bibfield  {author} {\bibinfo {author} {\bibfnamefont {A.}~\bibnamefont
  {Rezavand}}, \bibinfo {author} {\bibfnamefont {N.}~\bibnamefont {Ghobadi}}, \
  and\ \bibinfo {author} {\bibfnamefont {B.}~\bibnamefont {Behnamghader}},\
  }\href@noop {} {\bibfield  {journal} {\bibinfo  {journal} {Physical Review
  B}\ }\textbf {\bibinfo {volume} {106}},\ \bibinfo {pages} {035417} (\bibinfo
  {year} {2022})}\BibitemShut {NoStop}%
\bibitem [{\citenamefont {Guo}\ \emph {et~al.}(2021{\natexlab{b}})\citenamefont
  {Guo}, \citenamefont {Mu}, \citenamefont {Zhu}, \citenamefont {Han},\ and\
  \citenamefont {Ren}}]{guo2021predicted}%
  \BibitemOpen
  \bibfield  {author} {\bibinfo {author} {\bibfnamefont {S.-D.}\ \bibnamefont
  {Guo}}, \bibinfo {author} {\bibfnamefont {W.-Q.}\ \bibnamefont {Mu}},
  \bibinfo {author} {\bibfnamefont {Y.-T.}\ \bibnamefont {Zhu}}, \bibinfo
  {author} {\bibfnamefont {R.-Y.}\ \bibnamefont {Han}}, \ and\ \bibinfo
  {author} {\bibfnamefont {W.-C.}\ \bibnamefont {Ren}},\ }\href@noop {}
  {\bibfield  {journal} {\bibinfo  {journal} {Journal of Materials Chemistry
  C}\ }\textbf {\bibinfo {volume} {9}},\ \bibinfo {pages} {2464} (\bibinfo
  {year} {2021}{\natexlab{b}})}\BibitemShut {NoStop}%
\bibitem [{\citenamefont {Hussain}\ \emph {et~al.}(2022)\citenamefont
  {Hussain}, \citenamefont {Samad}, \citenamefont {Rehman}, \citenamefont
  {Cuono},\ and\ \citenamefont {Autieri}}]{hussain2022emergence}%
  \BibitemOpen
  \bibfield  {author} {\bibinfo {author} {\bibfnamefont {G.}~\bibnamefont
  {Hussain}}, \bibinfo {author} {\bibfnamefont {A.}~\bibnamefont {Samad}},
  \bibinfo {author} {\bibfnamefont {M.~U.}\ \bibnamefont {Rehman}}, \bibinfo
  {author} {\bibfnamefont {G.}~\bibnamefont {Cuono}}, \ and\ \bibinfo {author}
  {\bibfnamefont {C.}~\bibnamefont {Autieri}},\ }\href@noop {} {\bibfield
  {journal} {\bibinfo  {journal} {Journal of Magnetism and Magnetic Materials}\
  ,\ \bibinfo {pages} {169897}} (\bibinfo {year} {2022})}\BibitemShut {NoStop}%
\bibitem [{\citenamefont {Dey}\ \emph {et~al.}(2022)\citenamefont {Dey},
  \citenamefont {Ray},\ and\ \citenamefont {Yu}}]{dey2022intrinsic}%
  \BibitemOpen
  \bibfield  {author} {\bibinfo {author} {\bibfnamefont {D.}~\bibnamefont
  {Dey}}, \bibinfo {author} {\bibfnamefont {A.}~\bibnamefont {Ray}}, \ and\
  \bibinfo {author} {\bibfnamefont {L.}~\bibnamefont {Yu}},\ }\href@noop {}
  {\bibfield  {journal} {\bibinfo  {journal} {Physical Review Materials}\
  }\textbf {\bibinfo {volume} {6}},\ \bibinfo {pages} {L061002} (\bibinfo
  {year} {2022})}\BibitemShut {NoStop}%
\bibitem [{\citenamefont {Guo}\ \emph {et~al.}(2022)\citenamefont {Guo},
  \citenamefont {Mu}, \citenamefont {Wang}, \citenamefont {Yang}, \citenamefont
  {Wang},\ and\ \citenamefont {Ang}}]{guo2022strain}%
  \BibitemOpen
  \bibfield  {author} {\bibinfo {author} {\bibfnamefont {S.-D.}\ \bibnamefont
  {Guo}}, \bibinfo {author} {\bibfnamefont {W.-Q.}\ \bibnamefont {Mu}},
  \bibinfo {author} {\bibfnamefont {J.-H.}\ \bibnamefont {Wang}}, \bibinfo
  {author} {\bibfnamefont {Y.-X.}\ \bibnamefont {Yang}}, \bibinfo {author}
  {\bibfnamefont {B.}~\bibnamefont {Wang}}, \ and\ \bibinfo {author}
  {\bibfnamefont {Y.-S.}\ \bibnamefont {Ang}},\ }\href@noop {} {\bibfield
  {journal} {\bibinfo  {journal} {Physical Review B}\ }\textbf {\bibinfo
  {volume} {106}},\ \bibinfo {pages} {064416} (\bibinfo {year}
  {2022})}\BibitemShut {NoStop}%
\bibitem [{\citenamefont {Xiao}\ \emph {et~al.}(2012)\citenamefont {Xiao},
  \citenamefont {Liu}, \citenamefont {Feng}, \citenamefont {Xu},\ and\
  \citenamefont {Yao}}]{xiao2012coupled}%
  \BibitemOpen
  \bibfield  {author} {\bibinfo {author} {\bibfnamefont {D.}~\bibnamefont
  {Xiao}}, \bibinfo {author} {\bibfnamefont {G.-B.}\ \bibnamefont {Liu}},
  \bibinfo {author} {\bibfnamefont {W.}~\bibnamefont {Feng}}, \bibinfo {author}
  {\bibfnamefont {X.}~\bibnamefont {Xu}}, \ and\ \bibinfo {author}
  {\bibfnamefont {W.}~\bibnamefont {Yao}},\ }\href@noop {} {\bibfield
  {journal} {\bibinfo  {journal} {Phys. Rev. Lett.}\ }\textbf {\bibinfo
  {volume} {108}},\ \bibinfo {pages} {196802} (\bibinfo {year}
  {2012})}\BibitemShut {NoStop}%
\bibitem [{\citenamefont {Islam}\ \emph {et~al.}(2021)\citenamefont {Islam},
  \citenamefont {Ghosh}, \citenamefont {Autieri}, \citenamefont {Chowdhury},
  \citenamefont {Bansil}, \citenamefont {Agarwal},\ and\ \citenamefont
  {Singh}}]{islam2021tunable}%
  \BibitemOpen
  \bibfield  {author} {\bibinfo {author} {\bibfnamefont {R.}~\bibnamefont
  {Islam}}, \bibinfo {author} {\bibfnamefont {B.}~\bibnamefont {Ghosh}},
  \bibinfo {author} {\bibfnamefont {C.}~\bibnamefont {Autieri}}, \bibinfo
  {author} {\bibfnamefont {S.}~\bibnamefont {Chowdhury}}, \bibinfo {author}
  {\bibfnamefont {A.}~\bibnamefont {Bansil}}, \bibinfo {author} {\bibfnamefont
  {A.}~\bibnamefont {Agarwal}}, \ and\ \bibinfo {author} {\bibfnamefont
  {B.}~\bibnamefont {Singh}},\ }\href@noop {} {\bibfield  {journal} {\bibinfo
  {journal} {Phys. Rev. B}\ }\textbf {\bibinfo {volume} {104}},\ \bibinfo
  {pages} {L201112} (\bibinfo {year} {2021})}\BibitemShut {NoStop}%
\bibitem [{\citenamefont {Battilomo}\ \emph {et~al.}(2019)\citenamefont
  {Battilomo}, \citenamefont {Scopigno},\ and\ \citenamefont
  {Ortix}}]{PhysRevLett.123.196403}%
  \BibitemOpen
  \bibfield  {author} {\bibinfo {author} {\bibfnamefont {R.}~\bibnamefont
  {Battilomo}}, \bibinfo {author} {\bibfnamefont {N.}~\bibnamefont {Scopigno}},
  \ and\ \bibinfo {author} {\bibfnamefont {C.}~\bibnamefont {Ortix}},\ }\href
  {\doibase 10.1103/PhysRevLett.123.196403} {\bibfield  {journal} {\bibinfo
  {journal} {Phys. Rev. Lett.}\ }\textbf {\bibinfo {volume} {123}},\ \bibinfo
  {pages} {196403} (\bibinfo {year} {2019})}\BibitemShut {NoStop}%
\bibitem [{\citenamefont {Hurand}\ \emph {et~al.}(2015)\citenamefont {Hurand},
  \citenamefont {Jouan}, \citenamefont {Feuillet-Palma}, \citenamefont {Singh},
  \citenamefont {Biscaras}, \citenamefont {Lesne}, \citenamefont {Reyren},
  \citenamefont {Barth{\'e}l{\'e}my}, \citenamefont {Bibes}, \citenamefont
  {Villegas} \emph {et~al.}}]{hurand2015field}%
  \BibitemOpen
  \bibfield  {author} {\bibinfo {author} {\bibfnamefont {S.}~\bibnamefont
  {Hurand}}, \bibinfo {author} {\bibfnamefont {A.}~\bibnamefont {Jouan}},
  \bibinfo {author} {\bibfnamefont {C.}~\bibnamefont {Feuillet-Palma}},
  \bibinfo {author} {\bibfnamefont {G.}~\bibnamefont {Singh}}, \bibinfo
  {author} {\bibfnamefont {J.}~\bibnamefont {Biscaras}}, \bibinfo {author}
  {\bibfnamefont {E.}~\bibnamefont {Lesne}}, \bibinfo {author} {\bibfnamefont
  {N.}~\bibnamefont {Reyren}}, \bibinfo {author} {\bibfnamefont
  {A.}~\bibnamefont {Barth{\'e}l{\'e}my}}, \bibinfo {author} {\bibfnamefont
  {M.}~\bibnamefont {Bibes}}, \bibinfo {author} {\bibfnamefont
  {J.}~\bibnamefont {Villegas}},  \emph {et~al.},\ }\href@noop {} {\bibfield
  {journal} {\bibinfo  {journal} {Scientific reports}\ }\textbf {\bibinfo
  {volume} {5}},\ \bibinfo {pages} {1} (\bibinfo {year} {2015})}\BibitemShut
  {NoStop}%
\bibitem [{\citenamefont {Sheoran}\ \emph
  {et~al.}(2022{\natexlab{a}})\citenamefont {Sheoran}, \citenamefont {Kumar},
  \citenamefont {Bhumla},\ and\ \citenamefont {Bhattacharya}}]{D1MA00912E}%
  \BibitemOpen
  \bibfield  {author} {\bibinfo {author} {\bibfnamefont {S.}~\bibnamefont
  {Sheoran}}, \bibinfo {author} {\bibfnamefont {M.}~\bibnamefont {Kumar}},
  \bibinfo {author} {\bibfnamefont {P.}~\bibnamefont {Bhumla}}, \ and\ \bibinfo
  {author} {\bibfnamefont {S.}~\bibnamefont {Bhattacharya}},\ }\href {\doibase
  10.1039/D1MA00912E} {\bibfield  {journal} {\bibinfo  {journal} {Mater. Adv.}\
  }\textbf {\bibinfo {volume} {3}},\ \bibinfo {pages} {4170} (\bibinfo {year}
  {2022}{\natexlab{a}})}\BibitemShut {NoStop}%
\bibitem [{\citenamefont {Sheoran}\ \emph
  {et~al.}(2022{\natexlab{b}})\citenamefont {Sheoran}, \citenamefont {Bhumla},\
  and\ \citenamefont {Bhattacharya}}]{PhysRevMaterials.6.094602}%
  \BibitemOpen
  \bibfield  {author} {\bibinfo {author} {\bibfnamefont {S.}~\bibnamefont
  {Sheoran}}, \bibinfo {author} {\bibfnamefont {P.}~\bibnamefont {Bhumla}}, \
  and\ \bibinfo {author} {\bibfnamefont {S.}~\bibnamefont {Bhattacharya}},\
  }\href {\doibase 10.1103/PhysRevMaterials.6.094602} {\bibfield  {journal}
  {\bibinfo  {journal} {Phys. Rev. Materials}\ }\textbf {\bibinfo {volume}
  {6}},\ \bibinfo {pages} {094602} (\bibinfo {year}
  {2022}{\natexlab{b}})}\BibitemShut {NoStop}%
\bibitem [{\citenamefont {Xiao}\ \emph {et~al.}(2007)\citenamefont {Xiao},
  \citenamefont {Yao},\ and\ \citenamefont {Niu}}]{xiao2007valley}%
  \BibitemOpen
  \bibfield  {author} {\bibinfo {author} {\bibfnamefont {D.}~\bibnamefont
  {Xiao}}, \bibinfo {author} {\bibfnamefont {W.}~\bibnamefont {Yao}}, \ and\
  \bibinfo {author} {\bibfnamefont {Q.}~\bibnamefont {Niu}},\ }\href@noop {}
  {\bibfield  {journal} {\bibinfo  {journal} {Physical review letters}\
  }\textbf {\bibinfo {volume} {99}},\ \bibinfo {pages} {236809} (\bibinfo
  {year} {2007})}\BibitemShut {NoStop}%
\bibitem [{\citenamefont {Ahammed}\ and\ \citenamefont
  {De~Sarkar}(2022)}]{ahammed2022valley}%
  \BibitemOpen
  \bibfield  {author} {\bibinfo {author} {\bibfnamefont {R.}~\bibnamefont
  {Ahammed}}\ and\ \bibinfo {author} {\bibfnamefont {A.}~\bibnamefont
  {De~Sarkar}},\ }\href@noop {} {\bibfield  {journal} {\bibinfo  {journal}
  {Physical Review B}\ }\textbf {\bibinfo {volume} {105}},\ \bibinfo {pages}
  {045426} (\bibinfo {year} {2022})}\BibitemShut {NoStop}%
\bibitem [{\citenamefont {Jin}\ \emph {et~al.}(2021)\citenamefont {Jin},
  \citenamefont {Oh}, \citenamefont {Stania}, \citenamefont {Liu},\ and\
  \citenamefont {Yeom}}]{jin2021enhanced}%
  \BibitemOpen
  \bibfield  {author} {\bibinfo {author} {\bibfnamefont {K.-H.}\ \bibnamefont
  {Jin}}, \bibinfo {author} {\bibfnamefont {E.}~\bibnamefont {Oh}}, \bibinfo
  {author} {\bibfnamefont {R.}~\bibnamefont {Stania}}, \bibinfo {author}
  {\bibfnamefont {F.}~\bibnamefont {Liu}}, \ and\ \bibinfo {author}
  {\bibfnamefont {H.~W.}\ \bibnamefont {Yeom}},\ }\href@noop {} {\bibfield
  {journal} {\bibinfo  {journal} {Nano Lett.}\ }\textbf {\bibinfo {volume}
  {21}},\ \bibinfo {pages} {9468} (\bibinfo {year} {2021})}\BibitemShut
  {NoStop}%
\bibitem [{\citenamefont {Kresse}\ and\ \citenamefont
  {Furthm{\"u}ller}(1996)}]{kresse1996efficient}%
  \BibitemOpen
  \bibfield  {author} {\bibinfo {author} {\bibfnamefont {G.}~\bibnamefont
  {Kresse}}\ and\ \bibinfo {author} {\bibfnamefont {J.}~\bibnamefont
  {Furthm{\"u}ller}},\ }\href@noop {} {\bibfield  {journal} {\bibinfo
  {journal} {Phys. Rev. B}\ }\textbf {\bibinfo {volume} {54}},\ \bibinfo
  {pages} {11169} (\bibinfo {year} {1996})}\BibitemShut {NoStop}%
\bibitem [{\citenamefont {Bl{\"o}chl}(1994)}]{blochl1994projector}%
  \BibitemOpen
  \bibfield  {author} {\bibinfo {author} {\bibfnamefont {P.~E.}\ \bibnamefont
  {Bl{\"o}chl}},\ }\href@noop {} {\bibfield  {journal} {\bibinfo  {journal}
  {Phys. Rev. B}\ }\textbf {\bibinfo {volume} {50}},\ \bibinfo {pages} {17953}
  (\bibinfo {year} {1994})}\BibitemShut {NoStop}%
\bibitem [{\citenamefont {Kresse}\ and\ \citenamefont
  {Joubert}(1999)}]{kresse1999ultrasoft}%
  \BibitemOpen
  \bibfield  {author} {\bibinfo {author} {\bibfnamefont {G.}~\bibnamefont
  {Kresse}}\ and\ \bibinfo {author} {\bibfnamefont {D.}~\bibnamefont
  {Joubert}},\ }\href@noop {} {\bibfield  {journal} {\bibinfo  {journal} {Phys.
  Rev. B}\ }\textbf {\bibinfo {volume} {59}},\ \bibinfo {pages} {1758}
  (\bibinfo {year} {1999})}\BibitemShut {NoStop}%
\bibitem [{\citenamefont {Perdew}\ \emph {et~al.}(1996)\citenamefont {Perdew},
  \citenamefont {Burke},\ and\ \citenamefont
  {Ernzerhof}}]{perdew1996generalized}%
  \BibitemOpen
  \bibfield  {author} {\bibinfo {author} {\bibfnamefont {J.~P.}\ \bibnamefont
  {Perdew}}, \bibinfo {author} {\bibfnamefont {K.}~\bibnamefont {Burke}}, \
  and\ \bibinfo {author} {\bibfnamefont {M.}~\bibnamefont {Ernzerhof}},\
  }\href@noop {} {\bibfield  {journal} {\bibinfo  {journal} {Phys. Rev. Lett.}\
  }\textbf {\bibinfo {volume} {77}},\ \bibinfo {pages} {3865} (\bibinfo {year}
  {1996})}\BibitemShut {NoStop}%
\bibitem [{\citenamefont {Togo}\ and\ \citenamefont
  {Tanaka}(2015)}]{togo2015first}%
  \BibitemOpen
  \bibfield  {author} {\bibinfo {author} {\bibfnamefont {A.}~\bibnamefont
  {Togo}}\ and\ \bibinfo {author} {\bibfnamefont {I.}~\bibnamefont {Tanaka}},\
  }\href@noop {} {\bibfield  {journal} {\bibinfo  {journal} {Scr. Mater.}\
  }\textbf {\bibinfo {volume} {108}},\ \bibinfo {pages} {1} (\bibinfo {year}
  {2015})}\BibitemShut {NoStop}%
\bibitem [{\citenamefont {Mostofi}\ \emph {et~al.}(2014)\citenamefont
  {Mostofi}, \citenamefont {Yates}, \citenamefont {Pizzi}, \citenamefont {Lee},
  \citenamefont {Souza}, \citenamefont {Vanderbilt},\ and\ \citenamefont
  {Marzari}}]{mostofi2014updated}%
  \BibitemOpen
  \bibfield  {author} {\bibinfo {author} {\bibfnamefont {A.~A.}\ \bibnamefont
  {Mostofi}}, \bibinfo {author} {\bibfnamefont {J.~R.}\ \bibnamefont {Yates}},
  \bibinfo {author} {\bibfnamefont {G.}~\bibnamefont {Pizzi}}, \bibinfo
  {author} {\bibfnamefont {Y.-S.}\ \bibnamefont {Lee}}, \bibinfo {author}
  {\bibfnamefont {I.}~\bibnamefont {Souza}}, \bibinfo {author} {\bibfnamefont
  {D.}~\bibnamefont {Vanderbilt}}, \ and\ \bibinfo {author} {\bibfnamefont
  {N.}~\bibnamefont {Marzari}},\ }\href@noop {} {\bibfield  {journal} {\bibinfo
   {journal} {Computer Physics Communications}\ }\textbf {\bibinfo {volume}
  {185}},\ \bibinfo {pages} {2309} (\bibinfo {year} {2014})}\BibitemShut
  {NoStop}%
\bibitem [{\citenamefont {Heyd}\ \emph {et~al.}(2003)\citenamefont {Heyd},
  \citenamefont {Scuseria},\ and\ \citenamefont {Ernzerhof}}]{heyd2003hybrid}%
  \BibitemOpen
  \bibfield  {author} {\bibinfo {author} {\bibfnamefont {J.}~\bibnamefont
  {Heyd}}, \bibinfo {author} {\bibfnamefont {G.~E.}\ \bibnamefont {Scuseria}},
  \ and\ \bibinfo {author} {\bibfnamefont {M.}~\bibnamefont {Ernzerhof}},\
  }\href@noop {} {\bibfield  {journal} {\bibinfo  {journal} {The Journal of
  chemical physics}\ }\textbf {\bibinfo {volume} {118}},\ \bibinfo {pages}
  {8207} (\bibinfo {year} {2003})}\BibitemShut {NoStop}%
\bibitem [{\citenamefont {Hedin}(1965)}]{hedin1965new}%
  \BibitemOpen
  \bibfield  {author} {\bibinfo {author} {\bibfnamefont {L.}~\bibnamefont
  {Hedin}},\ }\href@noop {} {\bibfield  {journal} {\bibinfo  {journal} {Phys.
  Rev.}\ }\textbf {\bibinfo {volume} {139}},\ \bibinfo {pages} {A796} (\bibinfo
  {year} {1965})}\BibitemShut {NoStop}%
\bibitem [{\citenamefont {Hybertsen}\ and\ \citenamefont
  {Louie}(1985)}]{hybertsen1985first}%
  \BibitemOpen
  \bibfield  {author} {\bibinfo {author} {\bibfnamefont {M.~S.}\ \bibnamefont
  {Hybertsen}}\ and\ \bibinfo {author} {\bibfnamefont {S.~G.}\ \bibnamefont
  {Louie}},\ }\href@noop {} {\bibfield  {journal} {\bibinfo  {journal} {Phys.
  Rev. Lett.}\ }\textbf {\bibinfo {volume} {55}},\ \bibinfo {pages} {1418}
  (\bibinfo {year} {1985})}\BibitemShut {NoStop}%
\bibitem [{\citenamefont {Albrecht}\ \emph {et~al.}(1998)\citenamefont
  {Albrecht}, \citenamefont {Reining}, \citenamefont {Del~Sole},\ and\
  \citenamefont {Onida}}]{albrecht1998ab}%
  \BibitemOpen
  \bibfield  {author} {\bibinfo {author} {\bibfnamefont {S.}~\bibnamefont
  {Albrecht}}, \bibinfo {author} {\bibfnamefont {L.}~\bibnamefont {Reining}},
  \bibinfo {author} {\bibfnamefont {R.}~\bibnamefont {Del~Sole}}, \ and\
  \bibinfo {author} {\bibfnamefont {G.}~\bibnamefont {Onida}},\ }\href@noop {}
  {\bibfield  {journal} {\bibinfo  {journal} {Phys. Rev. Lett.}\ }\textbf
  {\bibinfo {volume} {80}},\ \bibinfo {pages} {4510} (\bibinfo {year}
  {1998})}\BibitemShut {NoStop}%
\bibitem [{\citenamefont {Sheoran}\ \emph
  {et~al.}(2022{\natexlab{c}})\citenamefont {Sheoran}, \citenamefont {Gill},
  \citenamefont {Phutela},\ and\ \citenamefont
  {Bhattacharya}}]{sheoran2022coupled}%
  \BibitemOpen
  \bibfield  {author} {\bibinfo {author} {\bibfnamefont {S.}~\bibnamefont
  {Sheoran}}, \bibinfo {author} {\bibfnamefont {D.}~\bibnamefont {Gill}},
  \bibinfo {author} {\bibfnamefont {A.}~\bibnamefont {Phutela}}, \ and\
  \bibinfo {author} {\bibfnamefont {S.}~\bibnamefont {Bhattacharya}},\
  }\href@noop {} {\bibfield  {journal} {\bibinfo  {journal} {arXiv:2208.00127}\
  } (\bibinfo {year} {2022}{\natexlab{c}})}\BibitemShut {NoStop}%
\bibitem [{\citenamefont {Sheoran}\ \emph {et~al.}(2023)\citenamefont
  {Sheoran}, \citenamefont {Monga}, \citenamefont {Phutela},\ and\
  \citenamefont {Bhattacharya}}]{sheoran2023coupled}%
  \BibitemOpen
  \bibfield  {author} {\bibinfo {author} {\bibfnamefont {S.}~\bibnamefont
  {Sheoran}}, \bibinfo {author} {\bibfnamefont {S.}~\bibnamefont {Monga}},
  \bibinfo {author} {\bibfnamefont {A.}~\bibnamefont {Phutela}}, \ and\
  \bibinfo {author} {\bibfnamefont {S.}~\bibnamefont {Bhattacharya}},\
  }\href@noop {} {\bibfield  {journal} {\bibinfo  {journal} {J. Phys. Chem.
  Lett.}\ }\textbf {\bibinfo {volume} {14}},\ \bibinfo {pages} {1494} (\bibinfo
  {year} {2023})}\BibitemShut {NoStop}%
\bibitem [{\citenamefont {Zhou}\ \emph
  {et~al.}(2021{\natexlab{a}})\citenamefont {Zhou}, \citenamefont {Wu},
  \citenamefont {Li}, \citenamefont {Zhang},\ and\ \citenamefont
  {Ouyang}}]{zhou2021structural}%
  \BibitemOpen
  \bibfield  {author} {\bibinfo {author} {\bibfnamefont {W.}~\bibnamefont
  {Zhou}}, \bibinfo {author} {\bibfnamefont {L.}~\bibnamefont {Wu}}, \bibinfo
  {author} {\bibfnamefont {A.}~\bibnamefont {Li}}, \bibinfo {author}
  {\bibfnamefont {B.}~\bibnamefont {Zhang}}, \ and\ \bibinfo {author}
  {\bibfnamefont {F.}~\bibnamefont {Ouyang}},\ }\href@noop {} {\bibfield
  {journal} {\bibinfo  {journal} {The Journal of Physical Chemistry Letters}\
  }\textbf {\bibinfo {volume} {12}},\ \bibinfo {pages} {11622} (\bibinfo {year}
  {2021}{\natexlab{a}})}\BibitemShut {NoStop}%
\bibitem [{\citenamefont {Kang}\ and\ \citenamefont
  {Lin}(2021)}]{kang2021second}%
  \BibitemOpen
  \bibfield  {author} {\bibinfo {author} {\bibfnamefont {L.}~\bibnamefont
  {Kang}}\ and\ \bibinfo {author} {\bibfnamefont {Z.}~\bibnamefont {Lin}},\
  }\href@noop {} {\bibfield  {journal} {\bibinfo  {journal} {Physical Review
  B}\ }\textbf {\bibinfo {volume} {103}},\ \bibinfo {pages} {195404} (\bibinfo
  {year} {2021})}\BibitemShut {NoStop}%
\bibitem [{\citenamefont {Wang}\ \emph {et~al.}(2022)\citenamefont {Wang},
  \citenamefont {Jiang}, \citenamefont {Liu}, \citenamefont {Zhang},
  \citenamefont {Li}, \citenamefont {Liu}, \citenamefont {Sun}, \citenamefont
  {Weng},\ and\ \citenamefont {Chen}}]{wang2022two}%
  \BibitemOpen
  \bibfield  {author} {\bibinfo {author} {\bibfnamefont {L.}~\bibnamefont
  {Wang}}, \bibinfo {author} {\bibfnamefont {Y.}~\bibnamefont {Jiang}},
  \bibinfo {author} {\bibfnamefont {J.}~\bibnamefont {Liu}}, \bibinfo {author}
  {\bibfnamefont {S.}~\bibnamefont {Zhang}}, \bibinfo {author} {\bibfnamefont
  {J.}~\bibnamefont {Li}}, \bibinfo {author} {\bibfnamefont {P.}~\bibnamefont
  {Liu}}, \bibinfo {author} {\bibfnamefont {Y.}~\bibnamefont {Sun}}, \bibinfo
  {author} {\bibfnamefont {H.}~\bibnamefont {Weng}}, \ and\ \bibinfo {author}
  {\bibfnamefont {X.-Q.}\ \bibnamefont {Chen}},\ }\href@noop {} {\bibfield
  {journal} {\bibinfo  {journal} {Physical Review B}\ }\textbf {\bibinfo
  {volume} {106}},\ \bibinfo {pages} {155144} (\bibinfo {year}
  {2022})}\BibitemShut {NoStop}%
\bibitem [{\citenamefont {Ren}\ \emph {et~al.}(2022)\citenamefont {Ren},
  \citenamefont {Hu}, \citenamefont {Chen}, \citenamefont {Hu}, \citenamefont
  {Wang}, \citenamefont {Gong}, \citenamefont {Zhang}, \citenamefont {Huang},\
  and\ \citenamefont {Shi}}]{ren2022two}%
  \BibitemOpen
  \bibfield  {author} {\bibinfo {author} {\bibfnamefont {Y.-T.}\ \bibnamefont
  {Ren}}, \bibinfo {author} {\bibfnamefont {L.}~\bibnamefont {Hu}}, \bibinfo
  {author} {\bibfnamefont {Y.-T.}\ \bibnamefont {Chen}}, \bibinfo {author}
  {\bibfnamefont {Y.-J.}\ \bibnamefont {Hu}}, \bibinfo {author} {\bibfnamefont
  {J.-L.}\ \bibnamefont {Wang}}, \bibinfo {author} {\bibfnamefont {P.-L.}\
  \bibnamefont {Gong}}, \bibinfo {author} {\bibfnamefont {H.}~\bibnamefont
  {Zhang}}, \bibinfo {author} {\bibfnamefont {L.}~\bibnamefont {Huang}}, \ and\
  \bibinfo {author} {\bibfnamefont {X.-Q.}\ \bibnamefont {Shi}},\ }\href@noop
  {} {\bibfield  {journal} {\bibinfo  {journal} {Physical Review Materials}\
  }\textbf {\bibinfo {volume} {6}},\ \bibinfo {pages} {064006} (\bibinfo {year}
  {2022})}\BibitemShut {NoStop}%
\bibitem [{\citenamefont {Guo}\ \emph {et~al.}(2021{\natexlab{c}})\citenamefont
  {Guo}, \citenamefont {Zhu}, \citenamefont {Mu}, \citenamefont {Wang},\ and\
  \citenamefont {Chen}}]{guo2021structure}%
  \BibitemOpen
  \bibfield  {author} {\bibinfo {author} {\bibfnamefont {S.-D.}\ \bibnamefont
  {Guo}}, \bibinfo {author} {\bibfnamefont {Y.-T.}\ \bibnamefont {Zhu}},
  \bibinfo {author} {\bibfnamefont {W.-Q.}\ \bibnamefont {Mu}}, \bibinfo
  {author} {\bibfnamefont {L.}~\bibnamefont {Wang}}, \ and\ \bibinfo {author}
  {\bibfnamefont {X.-Q.}\ \bibnamefont {Chen}},\ }\href@noop {} {\bibfield
  {journal} {\bibinfo  {journal} {Computational Materials Science}\ }\textbf
  {\bibinfo {volume} {188}},\ \bibinfo {pages} {110223} (\bibinfo {year}
  {2021}{\natexlab{c}})}\BibitemShut {NoStop}%
\bibitem [{\citenamefont {King-Smith}\ and\ \citenamefont
  {Vanderbilt}(1993)}]{king1993theory}%
  \BibitemOpen
  \bibfield  {author} {\bibinfo {author} {\bibfnamefont {R.}~\bibnamefont
  {King-Smith}}\ and\ \bibinfo {author} {\bibfnamefont {D.}~\bibnamefont
  {Vanderbilt}},\ }\href@noop {} {\bibfield  {journal} {\bibinfo  {journal}
  {Physical Review B}\ }\textbf {\bibinfo {volume} {47}},\ \bibinfo {pages}
  {1651} (\bibinfo {year} {1993})}\BibitemShut {NoStop}%
\bibitem [{\citenamefont {Spaldin}(2012)}]{spaldin2012beginner}%
  \BibitemOpen
  \bibfield  {author} {\bibinfo {author} {\bibfnamefont {N.~A.}\ \bibnamefont
  {Spaldin}},\ }\href@noop {} {\bibfield  {journal} {\bibinfo  {journal}
  {Journal of Solid State Chemistry}\ }\textbf {\bibinfo {volume} {195}},\
  \bibinfo {pages} {2} (\bibinfo {year} {2012})}\BibitemShut {NoStop}%
\bibitem [{\citenamefont {Yang}\ \emph {et~al.}(2021)\citenamefont {Yang},
  \citenamefont {Song}, \citenamefont {Sun},\ and\ \citenamefont
  {Lu}}]{yang2021valley}%
  \BibitemOpen
  \bibfield  {author} {\bibinfo {author} {\bibfnamefont {C.}~\bibnamefont
  {Yang}}, \bibinfo {author} {\bibfnamefont {Z.}~\bibnamefont {Song}}, \bibinfo
  {author} {\bibfnamefont {X.}~\bibnamefont {Sun}}, \ and\ \bibinfo {author}
  {\bibfnamefont {J.}~\bibnamefont {Lu}},\ }\href@noop {} {\bibfield  {journal}
  {\bibinfo  {journal} {Phys. Rev. B}\ }\textbf {\bibinfo {volume} {103}},\
  \bibinfo {pages} {035308} (\bibinfo {year} {2021})}\BibitemShut {NoStop}%
\bibitem [{\citenamefont {Li}\ \emph {et~al.}(2020)\citenamefont {Li},
  \citenamefont {Wu}, \citenamefont {Feng}, \citenamefont {Guan}, \citenamefont
  {Feng}, \citenamefont {Yao},\ and\ \citenamefont {Yang}}]{li2020valley}%
  \BibitemOpen
  \bibfield  {author} {\bibinfo {author} {\bibfnamefont {S.}~\bibnamefont
  {Li}}, \bibinfo {author} {\bibfnamefont {W.}~\bibnamefont {Wu}}, \bibinfo
  {author} {\bibfnamefont {X.}~\bibnamefont {Feng}}, \bibinfo {author}
  {\bibfnamefont {S.}~\bibnamefont {Guan}}, \bibinfo {author} {\bibfnamefont
  {W.}~\bibnamefont {Feng}}, \bibinfo {author} {\bibfnamefont {Y.}~\bibnamefont
  {Yao}}, \ and\ \bibinfo {author} {\bibfnamefont {S.~A.}\ \bibnamefont
  {Yang}},\ }\href@noop {} {\bibfield  {journal} {\bibinfo  {journal} {Phys.
  Rev. B}\ }\textbf {\bibinfo {volume} {102}},\ \bibinfo {pages} {235435}
  (\bibinfo {year} {2020})}\BibitemShut {NoStop}%
\bibitem [{\citenamefont {Lee}\ and\ \citenamefont
  {Choi}(2012)}]{PhysRevB.86.045437}%
  \BibitemOpen
  \bibfield  {author} {\bibinfo {author} {\bibfnamefont {H.}~\bibnamefont
  {Lee}}\ and\ \bibinfo {author} {\bibfnamefont {H.~J.}\ \bibnamefont {Choi}},\
  }\href {\doibase 10.1103/PhysRevB.86.045437} {\bibfield  {journal} {\bibinfo
  {journal} {Phys. Rev. B}\ }\textbf {\bibinfo {volume} {86}},\ \bibinfo
  {pages} {045437} (\bibinfo {year} {2012})}\BibitemShut {NoStop}%
\bibitem [{\citenamefont {Tian}\ \emph {et~al.}(2021)\citenamefont {Tian},
  \citenamefont {Liu}, \citenamefont {Shen}, \citenamefont {Li}, \citenamefont
  {Zhou}, \citenamefont {Liu}, \citenamefont {Chen},\ and\ \citenamefont
  {Yu}}]{tian2021manipulating}%
  \BibitemOpen
  \bibfield  {author} {\bibinfo {author} {\bibfnamefont {D.}~\bibnamefont
  {Tian}}, \bibinfo {author} {\bibfnamefont {Z.}~\bibnamefont {Liu}}, \bibinfo
  {author} {\bibfnamefont {S.}~\bibnamefont {Shen}}, \bibinfo {author}
  {\bibfnamefont {Z.}~\bibnamefont {Li}}, \bibinfo {author} {\bibfnamefont
  {Y.}~\bibnamefont {Zhou}}, \bibinfo {author} {\bibfnamefont {H.}~\bibnamefont
  {Liu}}, \bibinfo {author} {\bibfnamefont {H.}~\bibnamefont {Chen}}, \ and\
  \bibinfo {author} {\bibfnamefont {P.}~\bibnamefont {Yu}},\ }\href@noop {}
  {\bibfield  {journal} {\bibinfo  {journal} {Proceedings of the National
  Academy of Sciences}\ }\textbf {\bibinfo {volume} {118}},\ \bibinfo {pages}
  {e2101946118} (\bibinfo {year} {2021})}\BibitemShut {NoStop}%
\bibitem [{\citenamefont {Zhou}\ \emph
  {et~al.}(2021{\natexlab{b}})\citenamefont {Zhou}, \citenamefont {Chen},
  \citenamefont {Zhang}, \citenamefont {Duan},\ and\ \citenamefont
  {Ouyang}}]{zhou2021manipulation}%
  \BibitemOpen
  \bibfield  {author} {\bibinfo {author} {\bibfnamefont {W.}~\bibnamefont
  {Zhou}}, \bibinfo {author} {\bibfnamefont {J.}~\bibnamefont {Chen}}, \bibinfo
  {author} {\bibfnamefont {B.}~\bibnamefont {Zhang}}, \bibinfo {author}
  {\bibfnamefont {H.}~\bibnamefont {Duan}}, \ and\ \bibinfo {author}
  {\bibfnamefont {F.}~\bibnamefont {Ouyang}},\ }\href@noop {} {\bibfield
  {journal} {\bibinfo  {journal} {Physical Review B}\ }\textbf {\bibinfo
  {volume} {103}},\ \bibinfo {pages} {195114} (\bibinfo {year}
  {2021}{\natexlab{b}})}\BibitemShut {NoStop}%
\bibitem [{\citenamefont {Son}\ \emph {et~al.}(2019)\citenamefont {Son},
  \citenamefont {Kim}, \citenamefont {Ahn}, \citenamefont {Lee},\ and\
  \citenamefont {Lee}}]{son2019strain}%
  \BibitemOpen
  \bibfield  {author} {\bibinfo {author} {\bibfnamefont {J.}~\bibnamefont
  {Son}}, \bibinfo {author} {\bibfnamefont {K.-H.}\ \bibnamefont {Kim}},
  \bibinfo {author} {\bibfnamefont {Y.}~\bibnamefont {Ahn}}, \bibinfo {author}
  {\bibfnamefont {H.-W.}\ \bibnamefont {Lee}}, \ and\ \bibinfo {author}
  {\bibfnamefont {J.}~\bibnamefont {Lee}},\ }\href@noop {} {\bibfield
  {journal} {\bibinfo  {journal} {Physical review letters}\ }\textbf {\bibinfo
  {volume} {123}},\ \bibinfo {pages} {036806} (\bibinfo {year}
  {2019})}\BibitemShut {NoStop}%
\bibitem [{\citenamefont {Yu}\ \emph {et~al.}(2021)\citenamefont {Yu},
  \citenamefont {Zhou}, \citenamefont {Zhang},\ and\ \citenamefont
  {Chang}}]{yu2021spin}%
  \BibitemOpen
  \bibfield  {author} {\bibinfo {author} {\bibfnamefont {S.-B.}\ \bibnamefont
  {Yu}}, \bibinfo {author} {\bibfnamefont {M.}~\bibnamefont {Zhou}}, \bibinfo
  {author} {\bibfnamefont {D.}~\bibnamefont {Zhang}}, \ and\ \bibinfo {author}
  {\bibfnamefont {K.}~\bibnamefont {Chang}},\ }\href@noop {} {\bibfield
  {journal} {\bibinfo  {journal} {Physical Review B}\ }\textbf {\bibinfo
  {volume} {104}},\ \bibinfo {pages} {075435} (\bibinfo {year}
  {2021})}\BibitemShut {NoStop}%
\bibitem [{\citenamefont {Li}\ \emph {et~al.}(2019)\citenamefont {Li},
  \citenamefont {Wei}, \citenamefont {Wang}, \citenamefont {Huang},
  \citenamefont {Dai},\ and\ \citenamefont {Jacob}}]{li2019intrinsic}%
  \BibitemOpen
  \bibfield  {author} {\bibinfo {author} {\bibfnamefont {F.}~\bibnamefont
  {Li}}, \bibinfo {author} {\bibfnamefont {W.}~\bibnamefont {Wei}}, \bibinfo
  {author} {\bibfnamefont {H.}~\bibnamefont {Wang}}, \bibinfo {author}
  {\bibfnamefont {B.}~\bibnamefont {Huang}}, \bibinfo {author} {\bibfnamefont
  {Y.}~\bibnamefont {Dai}}, \ and\ \bibinfo {author} {\bibfnamefont
  {T.}~\bibnamefont {Jacob}},\ }\href@noop {} {\bibfield  {journal} {\bibinfo
  {journal} {The journal of physical chemistry letters}\ }\textbf {\bibinfo
  {volume} {10}},\ \bibinfo {pages} {559} (\bibinfo {year} {2019})}\BibitemShut
  {NoStop}%
\bibitem [{\citenamefont {Bhumla}\ \emph {et~al.}(2021)\citenamefont {Bhumla},
  \citenamefont {Gill}, \citenamefont {Sheoran},\ and\ \citenamefont
  {Bhattacharya}}]{bhumla2021origin}%
  \BibitemOpen
  \bibfield  {author} {\bibinfo {author} {\bibfnamefont {P.}~\bibnamefont
  {Bhumla}}, \bibinfo {author} {\bibfnamefont {D.}~\bibnamefont {Gill}},
  \bibinfo {author} {\bibfnamefont {S.}~\bibnamefont {Sheoran}}, \ and\
  \bibinfo {author} {\bibfnamefont {S.}~\bibnamefont {Bhattacharya}},\
  }\href@noop {} {\bibfield  {journal} {\bibinfo  {journal} {J. Phys. Chem.
  Lett.}\ }\textbf {\bibinfo {volume} {12}},\ \bibinfo {pages} {9539} (\bibinfo
  {year} {2021})}\BibitemShut {NoStop}%
\bibitem [{\citenamefont {Cheng}\ \emph {et~al.}(2016)\citenamefont {Cheng},
  \citenamefont {Sun}, \citenamefont {Chen}, \citenamefont {Fu},\ and\
  \citenamefont {Meng}}]{cheng2016nonlinear}%
  \BibitemOpen
  \bibfield  {author} {\bibinfo {author} {\bibfnamefont {C.}~\bibnamefont
  {Cheng}}, \bibinfo {author} {\bibfnamefont {J.-T.}\ \bibnamefont {Sun}},
  \bibinfo {author} {\bibfnamefont {X.-R.}\ \bibnamefont {Chen}}, \bibinfo
  {author} {\bibfnamefont {H.-X.}\ \bibnamefont {Fu}}, \ and\ \bibinfo {author}
  {\bibfnamefont {S.}~\bibnamefont {Meng}},\ }\href@noop {} {\bibfield
  {journal} {\bibinfo  {journal} {Nanoscale}\ }\textbf {\bibinfo {volume}
  {8}},\ \bibinfo {pages} {17854} (\bibinfo {year} {2016})}\BibitemShut
  {NoStop}%
\bibitem [{\citenamefont {Yao}\ \emph {et~al.}(2017)\citenamefont {Yao},
  \citenamefont {Cai}, \citenamefont {Tong}, \citenamefont {Gong},
  \citenamefont {Wang}, \citenamefont {Wan}, \citenamefont {Duan},\ and\
  \citenamefont {Chu}}]{yao2017manipulation}%
  \BibitemOpen
  \bibfield  {author} {\bibinfo {author} {\bibfnamefont {Q.-F.}\ \bibnamefont
  {Yao}}, \bibinfo {author} {\bibfnamefont {J.}~\bibnamefont {Cai}}, \bibinfo
  {author} {\bibfnamefont {W.-Y.}\ \bibnamefont {Tong}}, \bibinfo {author}
  {\bibfnamefont {S.-J.}\ \bibnamefont {Gong}}, \bibinfo {author}
  {\bibfnamefont {J.-Q.}\ \bibnamefont {Wang}}, \bibinfo {author}
  {\bibfnamefont {X.}~\bibnamefont {Wan}}, \bibinfo {author} {\bibfnamefont
  {C.-G.}\ \bibnamefont {Duan}}, \ and\ \bibinfo {author} {\bibfnamefont
  {J.}~\bibnamefont {Chu}},\ }\href@noop {} {\bibfield  {journal} {\bibinfo
  {journal} {Physical review B}\ }\textbf {\bibinfo {volume} {95}},\ \bibinfo
  {pages} {165401} (\bibinfo {year} {2017})}\BibitemShut {NoStop}%
\bibitem [{\citenamefont {Priydarshi}\ \emph {et~al.}(2022)\citenamefont
  {Priydarshi}, \citenamefont {Chauhan}, \citenamefont {Bhowmick},\ and\
  \citenamefont {Agarwal}}]{priydarshi2022large}%
  \BibitemOpen
  \bibfield  {author} {\bibinfo {author} {\bibfnamefont {A.}~\bibnamefont
  {Priydarshi}}, \bibinfo {author} {\bibfnamefont {Y.~S.}\ \bibnamefont
  {Chauhan}}, \bibinfo {author} {\bibfnamefont {S.}~\bibnamefont {Bhowmick}}, \
  and\ \bibinfo {author} {\bibfnamefont {A.}~\bibnamefont {Agarwal}},\
  }\href@noop {} {\bibfield  {journal} {\bibinfo  {journal} {Nanoscale}\ }
  (\bibinfo {year} {2022})}\BibitemShut {NoStop}%
\bibitem [{\citenamefont {Bruzzone}\ and\ \citenamefont
  {Fiori}(2011)}]{bruzzone2011ab}%
  \BibitemOpen
  \bibfield  {author} {\bibinfo {author} {\bibfnamefont {S.}~\bibnamefont
  {Bruzzone}}\ and\ \bibinfo {author} {\bibfnamefont {G.}~\bibnamefont
  {Fiori}},\ }\href@noop {} {\bibfield  {journal} {\bibinfo  {journal} {Applied
  Physics Letters}\ }\textbf {\bibinfo {volume} {99}},\ \bibinfo {pages}
  {222108} (\bibinfo {year} {2011})}\BibitemShut {NoStop}%
\end{thebibliography}%

\end{document}